\documentclass[preprint,12pt]{elsarticle}
\usepackage{amsmath}
\usepackage{amsfonts}
\usepackage{bm}
\usepackage{lipsum}
\usepackage{enumerate}
\usepackage{graphicx}
\usepackage{amsfonts}

\usepackage{mathrsfs}
\usepackage{subfigure}
\usepackage{epstopdf}
\usepackage{url}
\usepackage{verbatim}
\usepackage{amssymb}
\usepackage{hyperref}
\usepackage{color}
\usepackage{listings}

\def\@begintheorem#1#2{\par\bgroup{\scshape #1\ #2. }\it\ignorespaces}
\def\@opargbegintheorem#1#2#3{\par\bgroup%
   {\scshape #1\ #2\ ({\upshape #3}). }\it\ignorespaces}
\def\@endtheorem{\egroup}

\if@onethmnum
  \newtheorem{theorem}{Theorem}
  \newtheorem{lemma}[theorem]{Lemma}
  \newtheorem{corollary}[theorem]{Corollary}
  \newtheorem{proposition}[theorem]{Proposition}
  \newtheorem{definition}[theorem]{Definition}
\newtheorem{example}[theorem]{Example}
\newtheorem{remark}[theorem]{Remark}
\newtheorem{homework}[theorem]{Homework}
\newtheorem{case}[theorem]{}

\else

\fi

\usepackage[T1]{fontenc}

 \journal{Research in the Mathematical Sciences}

\begin{document}

\begin{frontmatter}
\title{Enhanced diffusivity and skewness of a diffusing tracer in the presence of an oscillating wall
}

\author[1]{Lingyun Ding}
\ead{dingly@live.unc.edu}
\author[1]{Robert Hunt}
\ead{huntrl@live.unc.edu}
\author[1]{Richard M. McLaughlin \corref{mycorrespondingauthor}}
\cortext[mycorrespondingauthor]{Corresponding author}
\ead{rmm@email.unc.edu}
\author[1]{Hunter Woodie}
\ead{woodieh@live.unc.edu}
\address[1]{Department of Mathematics, University of North Carolina, Chapel Hill, NC, 27599, United States}

\date{March 14, 2021}
\begin{abstract}
We develop a theory of enhanced diffusivity and skewness of the longitudinal distribution of a diffusing tracer advected by a periodic time-varying shear flow in a straight channel. Although applicable to any type of solute and fluid flow, we restrict the examples of our theory to the tracer advected by flows which are induced by a periodically oscillating wall in a Newtonian fluid between two infinite parallel plates as well as flow in an infinitely long duct.  These wall motions produce the well-known Stokes layer shear solutions which are exact solutions of the Navier-Stokes equations.  With these, we first calculate the second Aris moment for all time and its long-time limiting effective diffusivity as a function of the geometrical parameters, frequency, viscosity, and diffusivity.  Using a new formalism based upon the Helmholtz operator we establish a new single series formula for the variance valid for all time.  We show that the viscous dominated limit results in a linear shear layer for which the effective diffusivity is bounded with upper bound $\kappa(1+A^2/(2L^2))$, where $\kappa$ is the tracer diffusivity, $A$ is the amplitude of oscillation, and $L$ is the gap thickness.  Alternatively, for finite viscosities, we show that the enhanced diffusion is unbounded, diverging in the high-frequency limit.  Non-dimensionalization and physical arguments are given to explain these striking differences.  Asymptotics for the high-frequency behavior as well as the low viscosity limit are computed.  We present a study of the effective diffusivity surface as a function of the non-dimensional parameters which shows how a maximum can exists for various parameter sweeps.  Physical experiments are performed in water using particle tracking velocimetry (PTV) to quantitatively measure the fluid flow.  Using fluorescein dye as the passive tracer, we document that the theory is quantitatively accurate.  Specifically, image analysis suggests that the distribution variance be measured using the full width at half maximum statistic which is robust to noise.  Further, we show that the scalar skewness is zero for linear shear flows at all times, whereas for the nonlinear Stokes layer, exact analysis shows that the skewness sign can be controlled through the phase of the oscillating wall.  Further, for single frequency wall modes, we establish that the long-time skewness decays at the faster rate of $t^{-3/2}$ as compared with steady shear scalar skewness which decays at rate $t^{-1/2}$.  These results are confirmed using Monte-Carlo simulations.

\end{abstract}

\begin{keyword}
  Passive scalar \sep Effective diffusion \sep Skewness \sep Taylor dispersion \sep Multiscale analysis \sep Channel flow
\MSC[2010]{82C70, 82C80, 34E13, 76R50}
\end{keyword}
\end{frontmatter}

\section{Introduction}
\label{intro}

An extremely important class of problems concerns how fluid motion can increase solute mixing. Since G. I. Taylor \cite{taylor1953dispersion} first introduced the calculation showing that a pressure driven flow in a pipe leads to a greatly enhanced effective diffusivity, the literature on this topic has exploded in many directions spanning many disciplines. The mathematics of this problem is particularly important and just one of the many areas of Modern Applied Mathematics which Andy Majda pioneered, starting with work on developing a rigorous formulation characterizing how a scale separated flow with general streamline topology can give rise to an effective diffusivity \cite{avellaneda1989stieltjes,majda1993effect}, extending to non-scale separated flows showing anomalous results  \cite{avellaneda1990mathematical,avellaneda1992renormalization,avellaneda1992superdiffusion}, and eventually yielding models of scalar intermittency \cite{majda1993random,mclaughlin1996explicit} which produced explicit models for the full probability density function (PDF) of a passive scalar advected by a random, white in time linear shear layer  \cite{bronski2000rigorous,bronski2000problem,vanden2001non,camassa2010exact,vanden2001non,camassa2008evolution}.

Shortly following G. I. Taylor, Aris \cite{aris1956dispersion} presented an alternative approach for shear layers yielding a hierarchy for the spatial moments of the scalar field.  More recent results about the steady shear flow have explored how geometry can be used to control these moments to seek different effective diffusivities \cite{stone2004engineering,ajdari2006hydrodynamic}, and even how geometry can be used to control how solute in pressure-driven flow can be delivered with either a sharp front or with a gradual build-up through a detailed study of the scalar skewness \cite{aminian2016boundaries,aminian2015squaring,aminian2018mass}.

In many practical applications, flows are unsteady and therefore typically generate different properties than their steady counterparts. The first investigation of the Taylor dispersion in time-dependent flow dates back to Aris \cite{aris1960dispersion}, who presented the study of a solute in pulsating flow through a circular tube. After that, a number of studies reported on cases involving a non-transient, single-frequency pulsating flow\cite{bowden1965horizontal,chatwin1975longitudinal,watson1983diffusion,mukherjee1988dispersion,jimenez1984contaminant}. Most of those studies focused on pressure-driven flow; fewer studies have addressed wall driven flows.  Numerical studies of the enhanced mixing induced by a single frequency Couette-Poiseuille flow are reported in \cite{bandyopadhyay1999contaminant,paul2008dispersion}, and recently a multiscale analysis for a single frequency Couette-flow yielded formulae for the enhanced diffusivity \cite{barik2019multi,barik2017transport}.  

Recently, Vedel and Bruus \cite{vedel2012transient,vedel2014time} explored the case of a time-dependent, multifrequency flow and developed formulas of effective diffusivity.
Our study develops the general theory for the enhanced diffusion and skewness for the case of an arbitrary, periodic time varying shear flow and then focuses upon the physically realizable flow induced by the oscillatory motion of a wall adjacent to a Newtonian fluid theoretically, computationally, and experimentally.  First, we non-dimensionalize the problem and identify the non-dimensional parameters. Next, we derive the solutions to the Navier-Stokes equations resulting from such a wall motion, known as Stokes' second problem \cite{drazin2006navier}.  We see that in the high viscosity limit, this flow results in a time-varying linear shear layer. In turn, we compute the effective diffusivity produced by this flow, implementing a new formulation based on the Helmholtz operator which yields a new single series formula for the scalar variance in contrast to the double series formula in the literature, e.g. \cite{vedel2012transient}.  We establish an upper bound for the case of a time-varying linear shear, showing the maximum possible diffusion is set by the amplitude of wall motion and the gap thickness of the parallel-plate channel and is independent of the frequency of motion.  
Alternatively, we demonstrate that, for finite viscosities, the effective diffusivity is unbounded in increasing frequency of the wall motion.  These results are validated with experiments performed using a wall driven by a programmable linear motor.  Particle tracking velocimetry shows that the experimental fluid motion is accurately predicted by the Stokes layer solutions.  Image analysis with different camera exposure times suggests that dye distribution variances can be accurately measured using the full width at half maximum statistic.  Experiments with fluorescein dye are carried out and compare favorably with the effective diffusion theory.

We additionally study how the more nonlinear Stokes layer solutions can yield greater effective diffusivities than the linear counterpart.  Moreover, we document that the nonlinear case (with finite but nonzero viscosity) generates a much larger vertical concentration gradient, which leads to enhanced vertical tracer concentration on transient timescales. Next, we prove that for the case of the time-varying linear shear layer, the scalar spatial skewness is zero for all time, while Monte-Carlo simulations for wall-driven flows show that at finite viscosities the skewness can be non-zero.  Short-time asymptotics akin to prior work \cite{aminian2015squaring}  are computed for the skewness and compared directly to the Monte-Carlo simulations.  Finally, we present a complete mathematical analysis of the skewness showing how its sign can be completely controlled by the phase of the wall motion and further demonstrates that for single-frequency wall motions, the skewness decays to zero as $t^{-3/2}$ for large time, faster than the familiar steady flow counterpart, which decays as $t^{-1/2}$.  

\section{Theoretical calculations}

\subsection{Governing equation and nondimensionalization}

\subsubsection{Stokes Layer}

\begin{figure}
  \centering
  \includegraphics[width=1.0\linewidth]{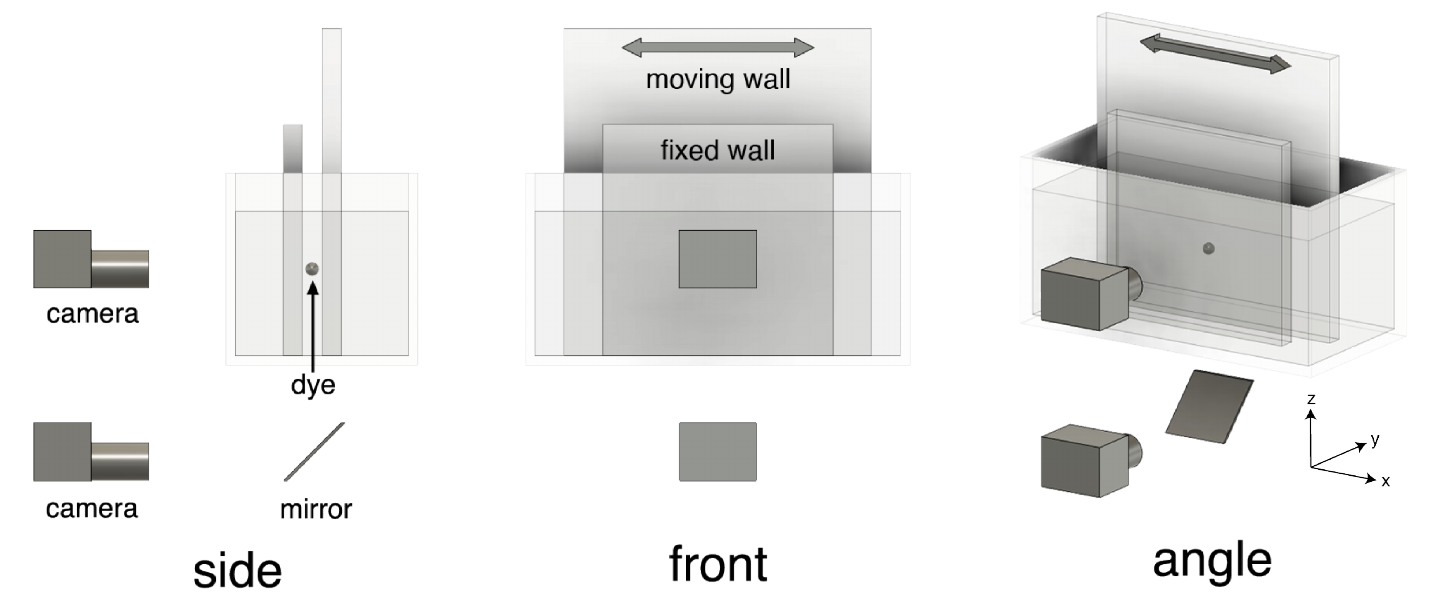}
  \hfill
  \caption[]
  {Schematic showing the setup for the experiment and theory.}
  \label{fig:schematic}
\end{figure}

We consider a layer of incompressible viscous fluid between two infinite parallel walls with gap thickness $L$. As sketched in the figure \ref{fig:schematic}, the front wall is stationary, while the back wall is moving periodically  parallel to itself with the velocity $\xi(t)$ and the base frequency  $\omega$. 
The flow $u(y,t)$ induced by the back  moving wall satisfies the Navier-Stokes equations:
\begin{equation}\label{eq:Stokes}
\partial_{t} u=  \nu \partial_{y}^2 u, \quad u(y,0)=0,\quad  u(0,t)=0,  \quad  u(L,t)=\xi( t),
\end{equation}
where $\nu$ is the fluid kinematic viscosity and the parallel-plate channel domain is  $\mathbb{R}\times \Omega$ and $x\in \mathbb{R}$, $\Omega = \left\{y  | y \in [0,L] \right\}$.  When $\xi (t)=A \omega \cos \omega t$, the long time solution of equation \eqref{eq:Stokes} is available in the chapter 4 of the book \cite{drazin2006navier} or equation 17 in the article \cite{mitran2008extensions}. This model was extended by Ferry and others to visco-elastic fluids \cite{ferry1947behavior,mitran2008extensions}. We derive the exact solution (with the transient term) and its high viscosity asymptotic expansion  in the appendix \ref{sec:TheStokesWaveinInfiniteChannel} for completeness. In three dimensional space, we are interested in the duct  $\mathbb{R} \times \Omega$, $\Omega = \left\{\left( y,z \right) | y \in [0,L], z\in [0,H] \right\}$. For the closed duct, the solid boundary imposes the no-slip boundary condition $ \left. u \right|_{z=0,H }=0$. For the open duct, we have the no-stress boundary condition at the free surface $\left. \frac{\partial u}{\partial z} \right|_{z=H }=0$. In both of these domains, for the parameters we used in our experiments, the analysis in appendix \ref{sec:TheStokeswaveDuct} shows the Stokes layer solution in a parallel-plate channel is a good approximation for the region away from the boundary in the $z$-direction. Hence, we neglect the boundary in the $z$-direction in the following calculation.

\subsubsection{ Advection-diffusion equation}
The passive scalar is governed by the advection-diffusion equation with a general time-varying shear flow $u(y,z,t)$ and no-flux boundary conditions which takes the form
\begin{equation}\label{eq:Advection Diffusion Equation deterministic shear}
  \partial_{t} T+u(y,z,t) \partial_{x} T =  \kappa \Delta T, \; T(x,y,z,0)= T_{I}(x,y,z),\;  \left. \partial_{\mathbf{n}} T  \right|_{\mathbb{R}\times \partial \Omega}=  0,
\end{equation}
where $\kappa$ is the diffusivity, $T_{I}(x,y,z)$ is the initial data and $\mathbf{n}$ is the outward normal vector of the boundary $\mathbb{R} \times \partial\Omega $.

\subsubsection{Nondimensionalization}
\label{sec:shearNondimensionalization}
With the change of variables
\begin{equation}
\begin{aligned}
&Lx'=x, \quad  Ly'=y, \quad Lz'=z, \quad\frac{L^2}{\kappa} t'=t, \quad \frac{\kappa}{L^2}\omega_{0}=\omega,\quad U=A\omega,\\
&  Uu' (y',z',t')= u (y,z,t), \quad U\xi' ( t')= \xi ( t),\quad L\Omega'=\Omega,\\
&T_I' (x',y',z') L^{-3}\int\limits_{\mathbb{R}\times \Omega}^{}T_{I}(x,y,z)\mathrm{d}x\mathrm{d}\Omega =T_{I} (x,y,z), \\
&T' (x',y',z',t') L^{-3}\int\limits_{\mathbb{R}\times \Omega}^{}T_{I} (x,y,z)\mathrm{d}x\mathrm{d}\Omega =T (x,y,z,t), \;
\end{aligned}
\end{equation}
after dropping the primes, we obtain the nondimensionalized flow equation
\begin{equation}\label{eq:StokesNon}
\partial_{t} u=  \mathrm{Sc} \partial_{y}^2 u,\quad u(y,0)=0, \quad  u(0,t)=0,  \quad  u(1,t)=\xi(t), 
\end{equation}
where $\mathrm{Sc}= {\nu}/{\kappa}$ is the Schmidt number. The dimensionless frequency $\omega_{0}$ also can be written as $\omega_{0}= \mathrm{Wo}^{2} \mathrm{Sc}$, where $\mathrm{Wo}=L\sqrt{\omega/ \nu}$ is the Womersley number. When $\xi (t)=\cos \omega_{0} t$, the long time solution of equation \eqref{eq:StokesNon} is
\begin{equation}\label{eq:StokeswaveVelocityNon}
\begin{aligned}
u(y,t)&=\sum\limits_{k=\pm 1}^{} \frac{\exp \left(\mathrm{i} t k\omega _0\right) \sinh \left(e^{\mathrm{i} \frac{\pi}{4}} \sqrt{k} \mathrm{Wo} y\right)}
{2\sinh \left(e^{\mathrm{i} \frac{\pi}{4}} \sqrt{k}\mathrm{Wo}\right)}.
\end{aligned}
\end{equation}
At a fixed time, the Womersley number uniquely determines the spatial shape of the Stokes layer solution. The advection-diffusion equation \eqref{eq:Advection Diffusion Equation deterministic shear} becomes
\begin{equation}\label{eq:Advection Diffusion Equation deterministic shearNon}
\partial_{t} T+ \mathrm{Pe}u(y,z,t) \partial_{x} T=  \Delta T,\; T(x,y,z,0)= T_{I}(x,y,z),\; \left. \partial_{\mathbf{n}} T \right|_{\mathbb{R} \times \partial \Omega}=  0, 
\end{equation}
where $\mathrm{Pe}=  {U L}/{ \kappa}= {A\omega L}/{ \kappa}$ is the P\'{e}clet number, the domain is $\mathbb{R}\times \Omega$ and $x\in \mathbb{R}$, $\Omega = \left\{y  | y \in [0,1] \right\}$ for the two dimensional problem,  $\mathbb{R}\times \Omega$ and $x\in \mathbb{R}$, $\Omega = \left\{ (y,z)  | y \in [0,1], z\in \mathbb{R} \right\}$  for the three dimensional problem.

\subsection{Aris moment hierarchy}

The $n$th Aris moment is defined by $T_n (y,z,t) = \int\limits_{-\infty}^{\infty} x^n T(x,y,z,t) \mathrm{d} x$. With the assumption $T(\pm \infty, y,z, t)=0$, the Aris moments satisfy the recursive relationship called the Aris equation,
\begin{equation}\label{eq:ArisMomentDef}
\begin{aligned}
&(\partial_t- \Delta)T_n=   n (n-1)T_{n-2}+ n  \mathrm{Pe}u(y,z,t) T_{n-1},\\
&T_n(y,z,0)= \int\limits_{-\infty}^{\infty} x^n T_{I}(x,y,z) \mathrm{d} x, \quad
\left. \partial_{\mathbf{n}} T \right|_{\mathbb{R} \times \partial \Omega}=  0, \\
\end{aligned}
\end{equation}
where $T_{n}=0$ if $n\leq -1$. The full moments of $T$ are then obtained though the cross-sectional average of the moments $ \bar{T}_n = \frac{1}{\left| \Omega\right|}  \int\limits_{\Omega}^{}T_n\mathrm{d}y\mathrm{d}z$, where $\Omega=\left\{ (y,z)| y\in [0,1], z\in \mathbb{R} \right\}$ is the cross section and $\left| \Omega \right|$ is the area of $\Omega$. In this following context, we use the overline to denote the cross sectional average. Applying the divergence theorem and boundary conditions gives
\begin{equation}\label{eq:averArisMomentDef}
\begin{aligned}
\frac{\mathrm{d}  \bar{T}_{n} }{\mathrm{d} t}= &  n(n-1)  \bar{T}_{n-2}+ n  \mathrm{Pe}  \overline{u(y,z,t)  T_{n-1}} ,\\
 \bar{T}_n (0)=&\frac{1}{\left| \Omega \right|} \int\limits_{\Omega}^{}\int\limits_{-\infty}^{\infty} x^n T_{I}(x,y,z) \mathrm{d} x \mathrm{d}y \mathrm{d}z.\\
\end{aligned}
\end{equation}
The multiscale analysis in appendix \ref{sec: MultiscaleAnalysis}  suggests that, assuming a scale separation in the initial data, the solution of equation \eqref{eq:Advection Diffusion Equation deterministic shear} can be approximated by a diffusion equation with an effective diffusion coefficient. Inspired by this observation, we study the longitudinal effective diffusivity through the cross sectional average $ \bar{T}$.
The effective longitudinal  diffusivity is defined as
\begin{equation}\label{eq:effectiveDiffusivityDefinition}
\kappa_{\mathrm{eff}}=   \lim\limits_{t\rightarrow \infty}\frac{\mathrm{Var} ( \bar{T})}{2t},
\end{equation}
where $\mathrm{Var} (\bar{T})= \bar{T}_2 - \bar{T}_1^2$ is the variance of the cross sectional average $ \bar{T}$. In this paper, we use $\kappa_{\mathrm{eff}}$ to denote the dimensional effective diffusivity computed by the dimensional Aris moment and  use the $\tilde{\kappa}_{\mathrm{eff}}=\kappa_{\mathrm{eff}}/\kappa$ to denote the non-dimensional effective diffusivity.

We are also interested in the symmetry properties of $\bar{T}$. Skewness is the lowest order integral measure of the asymmetry of a real-valued probability distribution, which is defined as
\begin{equation}\label{eq:SkewnessDefinition}
  \mathrm{S} (\bar{T})=   \frac{ \bar{T}_3- 3 \bar{T}_{2} \bar{T}_{1}+2 \bar{T}_{1}^{3}}
  {\left(\bar{T}_2 - \bar{T}_1^2\right)^{\frac{3}{2}}}.
\end{equation}
For a unimodal distribution, negative skewness commonly indicates that the distribution has the property median$>$mean while positive skewness indicates that the median$<$mean, see \cite{macgillivray1981mean,aminian2018mass} for sufficient conditions which guarantee this correlation. The information of shape provided by the skewness could improve the design of microfluidic flow injection analysis \cite{aminian2018mass,trojanowicz2016recent} and chromatographic separation \cite{blom2003chip}.

\subsection{Enhanced diffusivity and skewness induced by a general periodic time-varying flow}\label{sec:FrameDiffusivitySkewness}

In this section, we derive the formulae for the enhanced diffusivity and skewness induced by a general periodic time varying flow $u (y,t)$ which has the Fourier series representation
\begin{equation}
\begin{aligned}
u (y,t)&= \sum\limits_{k=-\infty}^{\infty} u_{k} e^{\mathrm{i} k \omega_0 t},
\end{aligned}
\end{equation}
where
$u_k=\frac{\omega_{0}}{2\pi}\int\limits_0^{2\pi/ \omega_0} u
(y,t)e^{-\mathrm{i} k \omega_{0} t} \mathrm{d} t$. Several
observations and assumptions can simplify our calculation.  Firstly,
we take $T(x,y,z,0)=\delta(x)$ as the initial data. Hence
$T_0(y,z,0)=1$ and $T_n(y,z,0)=0$ for $n\geq 1$ by the definition
\eqref{eq:ArisMomentDef}. Since the initial function and flow studied
here are independent of $z$, the three dimensional advection-diffusion
equation \eqref{eq:Advection Diffusion Equation deterministic shearNon} reduces to an
equation in two spatial dimensions. Secondly, to shorten the
expression, we denote $\phi_{0}=1$, $\lambda_{0}=0$ and 
$\phi_n= \sqrt{2} \cos n \pi y$, $\lambda_n=n^{2}\pi^{2}$, $ n\geq 1$ as
the eigenfunctions and eigenvalues of the Laplace operator in the
cross section of the parallel-plate channel. Those eigenfunctions form an orthogonal basis on the  cross section $\Omega$ with respect to the inner product  $\left\langle f,g \right\rangle= \int\limits_0^1 f g \mathrm{d} y$. Thirdly, the centered cross sectional average, e.g.,
variance and skewness, is invariant under the Galilean transformation
$\tilde{x}=x- \int\limits_0^{t}\int\limits_0^1 u (y,s) \mathrm{d}y
\mathrm{d} s$. We consider the problem in a frame of reference moving
with the spatial mean speed $\bar{u}$. Then the advection-diffusion
equation \eqref{eq:Advection Diffusion Equation deterministic shearNon} has the same form
but a new shear flow $\tilde{u}=u-\bar{u}$ with $\tilde{u}_{k}=u_{k}-\bar{u}_{k}$. Hence, $\bar{T}_1=0$ for
all time which simplifies the calculation of variance and skewness of
$\bar{T}$.

To compute the effective longitudinal  diffusivity, we need to compute the Aris moments $T_{0}$, $T_{1}$,  $\bar{T}_2 $ in turn.
When $n=0$, equation \eqref{eq:ArisMomentDef} becomes
\begin{equation}
\partial_{t} T_0- \partial_{y}^{2} T_{0}= 0, \quad T_0(y,0)= 1, \quad \left. \partial_{y} T_{0} \right|_{ y=0,1}= 0.
\end{equation}
The solution is $T_0=1$. When $n=1$, equation \eqref{eq:ArisMomentDef} is
\begin{equation}\label{eq:Aris moment 1}
\partial_{t} T_1- \partial_{y}^{2} T_{1}= \mathrm{Pe}\tilde{u}(y,t)T_0, \quad T_1(y,0)= 0, \quad \left. \partial_{y} T_{1} \right|_{ y=0,1}= 0.
\end{equation}
Then  $T_1$ has the  series representation which takes the form
\begin{equation}
\begin{aligned}
T_1&= \mathrm{Pe}  \sum\limits_{k_{1}=-\infty}^{\infty}\sum\limits_{n=1}^{\infty}\left\langle u_{k_{1}},\phi_{n} \right\rangle \phi_n \frac{e^{\mathrm{i} k_{1}  \omega _0 t}-e^{-t \lambda _n}}{\lambda _n+\mathrm{i} k_{1} \omega _0} \\
&=\mathrm{Pe}  \sum\limits_{k_{1}=-\infty}^{\infty}\left( Q_{k_{1}}^{(1)} e^{\mathrm{i} k_{1}  \omega _0 t}- \sum\limits_{n=1}^{\infty}\frac{\left\langle u_{k_{1}},\phi_{n} \right\rangle \phi_n e^{-t \lambda _n}}{\lambda _n+\mathrm{i} k_{1} \omega _0}  \right), \\
\end{aligned}
\end{equation}
where $Q_{k_{1}}^{(1)}=( -\Delta+\mathrm{i} k_{1} \omega_0)^{-1} \left( u_{k} - \bar{u}_{k} \right)$ and the inverse Helmholtz operator, $b (y) =\left( -\Delta+\lambda \right)^{-1}a (y)$,  solves
\begin{equation}
-\partial_{y}^2 b (y) + \lambda b (y)= a (y), \quad \left. \partial_{y} b \right|_{ y=0,1}= 0.
\end{equation}
We note that $b (y)$ has the integral representation
\begin{equation}\label{eq:HelmholtzSol1D}
\begin{aligned}
b (y)= &\frac{1}{\sqrt{\lambda }}\left( \frac{\cosh \left(\sqrt{\lambda } y\right) \int_0^1 a(s) \cosh \left(\sqrt{\lambda } (1-s)\right) \mathrm{d} s}{\sinh \left(\sqrt{\lambda } \right)}\right.\\
&\hspace{1cm} \left. -\int_0^y a(s) \sinh \left(\sqrt{\lambda } (y-s)\right) \mathrm{d} s \right), \quad \lambda \neq -\lambda_{n}. \\
&b(y)=-\int_{0}^y \int_{0}^{y_1}a(y_{2})\mathrm{d}y_2\mathrm{d}y_1+\int_0^{1}\int_{0}^y \int_{0}^{y_1}a(y_{2})\mathrm{d}y_2\mathrm{d}y_1\mathrm{d}y, \quad \lambda=0.
\end{aligned}
\end{equation}
When $\lambda=-\lambda_n=-n^{2}\pi^{2}$, $a (y)$ should satisfy the solvability condition $\left\langle a, \phi_n \right\rangle=0$. In this case, the boundary value problem has infinite solutions.  We choose the particular solution which satisfies $\left\langle b,\gamma_n \right\rangle=0$,
\begin{equation}
\begin{aligned}
b (y)& =\left( -\Delta-\lambda_n \right)^{-1}a  =\lim\limits_{\lambda\rightarrow -\lambda_{n}}\left( -\Delta+\lambda \right)^{-1}a - \left\langle \left( -\Delta+\lambda \right)^{-1}a , \phi_{n} \right\rangle. \\
\end{aligned}
\end{equation}   When the cross section $\Omega$ has more general geometry, $b=(-\Delta+ \lambda)^{-1}a$ becomes the solution of the Helmholtz equation $(-\Delta+ \lambda)b=a$ on $\Omega$ with no-flux boundary conditions. 


When $n=2$, equation \eqref{eq:averArisMomentDef} is:
\begin{equation}
\frac{\mathrm{d}  \bar{T}_{2}}{\mathrm{d} t}= 2  \bar{T}_{0} + 2 \mathrm{Pe}\overline{\tilde{u}(y,t)  T_{1} },\quad \bar{T}_2(0)=0.
\end{equation}
By solving this equation, we have
\begin{equation}\label{eq:arisT2Full}
\begin{aligned}
  \bar{T}_2 =& 2t+ 2 \mathrm{Pe}^2  \sum\limits_{k_{1},k_{2}\in \mathbb{Z}}^{}\sum\limits_{n=1}^{\infty}\left\langle u_{k_{1}},\phi_n \right\rangle \left\langle u_{k_{2}},\phi_{n} \right\rangle  \left\{ \frac{-1+e^{\mathrm{i} \left(k_1+k_2\right) \omega _0 t} }{\left(k_1+k_2\right) \omega _0 \left(\mathrm{i} \lambda _n-k_1 \omega _0\right)} \right.\\
  &\hspace{5cm} \left. -\frac{1-e^{- \lambda _n t+\mathrm{i} k_2  \omega _0 t}}{\left(\lambda _n+\mathrm{i} k_1 \omega _0\right) \left(\lambda _n-\mathrm{i} k_2 \omega _0\right)}\right\},
\end{aligned}
\end{equation}
where the summand is understood as an entire function whose value is determined by its power series. For example, $f(z)=\frac{e^{z}-1}{z}=1+\frac{z}{2}+\mathcal{O}(z^{2})$, so $f(0)=1$.
The effective longitudinal diffusivity  defined in \eqref{eq:effectiveDiffusivityDefinition} is then
\begin{equation}\label{effectiveDiffusivityPeriodic}
\begin{aligned}
\tilde{\kappa}_{\mathrm{eff}}=& 1+  \mathrm{Pe}^2  \sum\limits_{k=- \infty}^{\infty}\sum\limits_{n=1}^{\infty}  \frac{\left\langle u_{k},\phi_n \right\rangle\left\langle u_{-k},\phi_n \right\rangle}{\lambda _n+\mathrm{i} k \omega _0}
=1+  \mathrm{Pe}^2 \sum\limits_{k=- \infty}^{\infty}  \left\langle Q_{k}^{(1)}, u_{-k} \right\rangle.
\end{aligned}
\end{equation}
The double series representation for effective diffusivity is identical to equation (3.24) in \cite{vedel2012transient}, while the single series representation presented here is new.  For the steady flow $\tilde{u} (y,t) =\tilde{u}_{0} (y)$, the last expression in equation \eqref{effectiveDiffusivityPeriodic} becomes equation (1.30) in  \cite{taylor2012random}. Moreover, by the divergence theorem, we have
\begin{equation}
\begin{aligned}
\tilde{\kappa}_{\mathrm{eff}}= & 1+ \mathrm{Pe}^{2}\left\langle -\tilde{u} \Delta^{-1}\tilde{u}  \right\rangle  =1+ \mathrm{Pe}^{2}\left\langle  \left( \nabla \Delta^{-1}\tilde{u} \right)\cdot  \left( \nabla \Delta^{-1}\tilde{u} \right)  \right\rangle=1+ \mathrm{Pe}^{2}\lVert \tilde{u}\rVert_{H^{-1}}, \\
\end{aligned}
\end{equation}
where $\lVert u \rVert_{H^{-1}}= \left\langle  \left( \nabla \Delta^{-1}u \right)\cdot  \left( \nabla \Delta^{-1}u \right)  \right\rangle$ is the $H^{-1}$ norm of $u$. Interestingly, the $H^{-1}$ norm is widely used for measuring mixing efficiency in the field of chaotic advection \cite{thiffeault2012using,lin2011optimal,lunasin2012optimal,aref2017frontiers}. It also appears in the effective diffusivity here which is a measurement of mixing efficiency in this shear dispersion problem.  

Comparing equation \eqref{eq:homogenization cell problem1 shear} and \eqref{eq:Aris moment 1}, we can see that the solution $\theta$ of the cell problem is the first Aris moment $T_{1}$. The formula of effective diffusivity \eqref{eq:steady homogenization effdiffusivity} is equivalent to equation \eqref{eq:effectiveDiffusivityDefinition}. Hence, we conclude that the Aris moment approach and the multiscale analysis approach yield the same effective diffusivity for the time-varying shear flow.  Of course, we note that the limiting procedure here, with $\epsilon \rightarrow 0$, may be different than the Aris moment approach where the limit is $t\rightarrow \infty$.

To compute the skewness of the cross sectional average $\bar{T}$, we need to compute the Aris moments $T_{2},$  $\bar{T}_3 $ in turn.
 When $n=2$, equation \eqref{eq:ArisMomentDef} is
\begin{equation}\label{eq:Aris moment 2}
\partial_{t} T_2- \partial_{y}^{2} T_{2}=2T_{0}+ 2\mathrm{Pe}\tilde{u}(y,t)T_1, \quad T_2(y,0)= 0, \quad \left. \partial_{y} T_{2} \right|_{ y=0,1}= 0.
\end{equation}
Here $\tilde{u}(y,t)T_1$ has the series representation 
\begin{equation}
\begin{aligned}
\tilde{u}(y,t)T_1=& \mathrm{Pe} \sum\limits_{k_{1},k_{2}\in \mathbb{Z}}^{} \sum\limits_{n_{1}=1,n_{2}=0}^{\infty} \left\langle u_{k_{1}},\phi_{n_{1}} \right\rangle \left\langle u_{k_2}\phi_{n_{1}},\phi_{n_{2}} \right\rangle  \phi_{n_{2}} \frac{e^{\mathrm{i} (k_{1}+k_{2})  \omega _0 t}-e^{\mathrm{i} k_{2}\omega_0 t- \lambda _{n_{1}} t}}{\lambda_{n_{1}}+\mathrm{i} k_{1} \omega _0}. \\
\end{aligned}
\end{equation}
Hence, $T_2$ has the series representation 
\begin{equation}
\begin{aligned}
T_2=&2t+ 2\mathrm{Pe}^{2} \sum\limits_{k_{1},k_{2}\in \mathbb{Z}}^{} \sum\limits_{n_{1}=1,n_{2}=0}^{\infty}
 \frac{\left\langle u_{k_{1}},\phi_{n_{1}} \right\rangle \left\langle u_{k_2}\phi_{n_{1}},\phi_{n_{2}} \right\rangle  \phi_{n_{2}} }{\lambda _{n_1}+\mathrm{i} k_1 \omega _0}\times
\\
& \left(
\frac{ -e^{- \lambda _{n_2} t}+e^{ \mathrm{i} \left(k_1+k_2\right) \omega _0 t}}{\lambda _{n_2}+\mathrm{i} \left(k_1+k_2\right) \omega _0}-\frac{e^{- \lambda _{n_2} t}-e^{- \lambda _{n_1}t+\mathrm{i} k_2  \omega _0 t}}{-\mathrm{i} k_2 \omega _0+\lambda _{n_1}-\lambda _{n_2}}
\right). \\
\end{aligned}
\end{equation}

When $n=3$, equation \eqref{eq:averArisMomentDef} is
\begin{equation}
\frac{\mathrm{d}  \bar{T}_{3}}{\mathrm{d} t}= 6  \bar{T}_{1} + 3 \mathrm{Pe}\overline{\tilde{u}(y,t)  T_{2} },\quad \bar{T}_3(0)=0.
\end{equation}
$\bar{T}_{1}=0$ follows from the choice of the frame of reference. Hence, we obtain
\begin{equation}\label{eq:ThirdArisMoment}
\begin{aligned}
 \bar{T}_{3}=&  6\mathrm{Pe}^{3}\sum\limits_{k_{1},k_{2},k_{3}\in \mathbb{Z}}^{} \sum\limits_{n_{1},n_{2}=1}^{\infty}
 \frac{ \left\langle u_{k_{1}},\phi_{n_{1}} \right\rangle\left\langle u_{k_2}\phi_{n_{1}},\phi_{n_{2}} \right\rangle \left\langle u_{k_{3}}, \phi_{n_{2}} \right\rangle }{\lambda _{n_1}+\mathrm{i} k_1 \omega _0}\times \\
 &\left\{ \tfrac{1}{-\mathrm{i} k_2 \omega _0+\lambda _{n_1}-\lambda _{n_2}} \left(
\tfrac{1-e^{-t \left(\lambda _{n_1}-\mathrm{i} \left(k_2+k_3\right) \omega _0\right)}}{\lambda _{n_1}-\mathrm{i} \left(k_2+k_3\right) \omega _0}-\tfrac{1-e^{-t \lambda _{n_2}+\mathrm{i} k_3 t \omega _0}}{\lambda _{n_2}-\mathrm{i} k_3 \omega _0}
   \right)
\right.
   \\
   &\left.-\tfrac{1}{\lambda _{n_2}+\mathrm{i} \left(k_1+k_2\right) \omega _0} \left(\tfrac{1-e^{-t \lambda _{n_2}+\mathrm{i} k_3 t \omega _0}}{\lambda _{n_2}-\mathrm{i} k_3 \omega _0}-\tfrac{-1+e^{\mathrm{i} \left(k_1+k_2+k_3\right) t \omega _0}}{\mathrm{i}\left(k_1+k_2+k_3\right) \omega _0}
     \right)
   \right\}.
\end{aligned}
\end{equation}
With the definition of skewness \eqref{eq:SkewnessDefinition} and $\bar{T}_1=0$, we have
\begin{equation}\label{eq:skewnessMultiF}
\begin{aligned}
  S (\bar{T})=& \tfrac{3\mathrm{Pe}^{3}}{\sqrt{ 2\tilde{\kappa}_{\mathrm{eff}}^{3} t}}\sum\limits_{k_{1},k_{2}\in \mathbb{Z}}^{} \sum\limits_{n_{1},n_{2}=1}^{\infty}  \tfrac{\left\langle u_{k_{1}},\phi_{n_{1}} \right\rangle \left\langle u_{k_2}\phi_{n_{1}},\phi_{n_{2}} \right\rangle \left\langle u_{-k_{1}-k_{2}}, \phi_{n_{2}}  \right\rangle}{\left(\lambda _{n_1}+\mathrm{i} k_1 \omega _0\right) \left(\lambda _{n_2}+\mathrm{i} \left(k_1+k_2\right) \omega _0\right)}  +\mathcal{O} (t^{-\frac{3}{2}})\\
=&\tfrac{3\mathrm{Pe}^{3} \sum\limits_{k_{1},k_{2} \in \mathbb{Z}}^{}  \left\langle Q^{(2,1)}_{k_{1},k_{2}}, u_{-k_{1}-k_{2}} \right\rangle}{ \sqrt{2t} \left( 1+  \mathrm{Pe}^2 \sum\limits_{k=- \infty}^{\infty}  \left( Q_{k}^{(1)}, u_{-k} \right\rangle \right)^{\frac{3}{2}}}
 +\mathcal{O} (t^{-\frac{3}{2}}).
\end{aligned}
\end{equation}
where $Q^{(2,1)}_{k_{1},k_{2}}=(\mathrm{i} (k_{1}+k_{2})\omega_{0}-\Delta)^{-1} \left( Q^{(1)}_{k_{1}}\tilde{u}_{k_{2}} - \overline{Q^{(1)}_{k_{1}}\tilde{u}_{k_{2}} } \right)$. For steady flow, equation \eqref{eq:skewnessMultiF} reduces to equation (24) in the supplementary materials of article \cite{aminian2016boundaries}. For a single frequency flow, $k_{i} \in \left\{ -1,1 \right\}$ and $\delta_{k_{1}+k_{2},-k_{3}}=0$ for all combinations of $k_{1},k_{2},k_{3}$. Hence, the leading order in equation \eqref{eq:skewnessMultiF} vanishes, which leads to the long time asymptotic expansion of skewness
\begin{equation}\label{eq:skewnessSingleF}
\begin{aligned}
  S (\bar{T})= & \tfrac{3\mathrm{Pe}^{3}}{\sqrt{ 2\tilde{\kappa}_{\mathrm{eff}}^{3} t^{3}}}\sum\limits_{k_{1},k_{2},k_{3}\in \mathbb{Z}}^{} \sum\limits_{n_{1},n_{2}=1}^{\infty}
 \frac{ \left\langle u_{k_{1}},\phi_{n_{1}} \right\rangle\left\langle u_{k_2}\phi_{n_{1}},\phi_{n_{2}} \right\rangle \left\langle u_{k_{3}}, \phi_{n_{2}} \right\rangle }{\lambda _{n_1}+\mathrm{i} k_1 \omega _0}\times \\
 &\left\{ \tfrac{1}{-\mathrm{i} k_2 \omega _0+\lambda _{n_1}-\lambda _{n_2}} \left(
\tfrac{1}{\lambda _{n_1}-\mathrm{i} \left(k_2+k_3\right) \omega _0}-\tfrac{1}{\lambda _{n_2}-\mathrm{i} k_3 \omega _0}
   \right)
\right.
   \\
   &\left.-\tfrac{1}{\lambda _{n_2}+\mathrm{i} \left(k_1+k_2\right) \omega _0} \left(\tfrac{1}{\lambda _{n_2}-\mathrm{i} k_3 \omega _0}-\tfrac{-1+e^{\mathrm{i} \left(k_1+k_2+k_3\right) t \omega _0}}{\mathrm{i}\left(k_1+k_2+k_3\right) \omega _0}
     \right)
   \right\}
 +\mathcal{O} (e^{-\lambda_{1}t}).\\
\end{aligned}
\end{equation}
This expression implies that single frequency flows or multiple frequency flows with suitable frequency separation could relax more quickly to a symmetric $\bar{T}$ than other flows, e.g., steady Poiseuille flow.

\subsection{Enhanced diffusivity induced by an oscillating wall}
With the formula we derived in the previous section, we present a detailed analysis of the enhanced diffusivity induced by the Stokes layer solution and its dependence on the parameters. With the formulae for the Stokes layer solution \eqref{eq:StokeswaveVelocityNon} and second Aris moment \eqref{eq:arisT2Full}, we have 
\begin{equation}\label{eq:SecondMomentStokes}
\begin{aligned}
  \bar{T}_2 =& 2t+ 2 \mathrm{Pe}^2  \sum\limits_{k_{1},k_{2}= \pm 1}^{}\sum\limits_{n=1}^{\infty} \left( \tfrac{-1+e^{\mathrm{i} \left(k_1+k_2\right) \omega _0 t} }{\left(k_1+k_2\right) \omega _0 \left(\mathrm{i} \lambda _n-k_1 \omega _0\right)} -\tfrac{1-e^{- \lambda _n t+\mathrm{i} k_2  \omega _0 t}}{\left(\lambda _n+\mathrm{i} k_1 \omega _0\right) \left(\lambda _n-\mathrm{i} k_2 \omega _0\right)}\right)\\
  &\times \prod\limits_{j=1}^2\tfrac{e^{\mathrm{i} \frac{\pi}{4}} \sqrt{k_{j}} \mathrm{Wo}  \left((-1)^n \cosh \left(e^{\mathrm{i} \frac{\pi}{4}} \sqrt{k_j} \mathrm{Wo}\right)-1\right)}
  {\sqrt{2}\mathrm{sinh}\left(e^{\mathrm{i} \frac{\pi}{4}} \sqrt{k_j} \mathrm{Wo}\right)\left( \pi ^2 n^2+\mathrm{i} k_j \mathrm{Wo}^2 \right)}. 
\end{aligned}
\end{equation}

With equation  \eqref{effectiveDiffusivityPeriodic}, the effective longitudinal diffusivity  induced by the Stokes layer solution  is  then
\begin{equation}\label{eq:ekappaStokes}
\begin{aligned}
\tilde{\kappa}_{\mathrm{eff}}=& 1+\frac{  \mathrm{Pe}^2 \mathrm{Wo}^2}{2 \sqrt{2}\left( \cosh (\sqrt{2} \mathrm{Wo})-\cos (\sqrt{2} \mathrm{Wo}) \right)}
\left\{-\frac{\sin (\sqrt{2} \mathrm{Wo})+\sinh (\sqrt{2} \mathrm{Wo})}{\mathrm{Wo}\left( \mathrm{Wo}^4- \omega_{0} ^2\right)}
\right. \\
  &+\frac{1}{\sqrt{\omega_{0} } \left(\omega_{0} ^2-\mathrm{Wo}^{4}\right) \left(\cos \left(\sqrt{2\omega_{0} }\right)-\cosh \left(\sqrt{2\omega_{0} }\right)\right)}\times\\
  &\left( 4 \sqrt{2} e^{\frac{\pi}{4}\mathrm{i} } \cos \left( \frac{\mathrm{Wo}}{\sqrt{2}} \right) \cosh \left( \frac{\mathrm{Wo}}{\sqrt{2}} \right) \left(\sin \left(e^{\frac{\pi}{4}\mathrm{i} } \sqrt{\omega_{0} }\right)-\sinh \left(e^{\frac{\pi}{4}\mathrm{i} } \sqrt{\omega_{0} }\right)\right) \right.\\
  &\left.\left. -(\cos (\sqrt{2} \mathrm{Wo})+\cosh (\sqrt{2} \mathrm{Wo})+2) \left(\sin \left(\sqrt{2\omega_{0} }\right)-\sinh \left(\sqrt{2\omega_{0} }\right)\right) \right)\right\}.\\
\end{aligned}
\end{equation}

The three non-dimensional parameters $\omega_0, \mathrm{Wo}, \mathrm{Sc}$ are connected by the relation $\omega_0= \mathrm{Wo}^{2} \mathrm{Sc}$. To study limiting cases, we need to assume two of them are independent and eliminate the remaining parameter from equation \eqref{eq:ekappaStokes}. We first study the low and high limit of Womersley number with a given $\omega_0$, i.e. $\mathrm{Sc}$ becomes a function of $\mathrm{Wo}$.  The expansion \eqref{eq:lowWoExpansion} shows that the Stokes layer solution converges to the linear shear flow  $u(y,t)= y \cos \left( \omega_{0} t \right)$ as $ \mathrm{Wo} \rightarrow 0$. In the low Womersley number limit, the effective diffusivity \eqref{eq:ekappaStokes} becomes
\begin{equation}\label{eq:effdiffusivityShear}
\begin{aligned}
\tilde{\kappa}_{\mathrm{eff}}=1+\frac{\mathrm{Pe}^2}{2 \omega_{0} ^2}  \left( 1-\frac{\sqrt{2}}{\sqrt{\omega_{0}}}\frac{\sin \left(\frac{\sqrt{\omega_{0} }}{\sqrt{2}}\right)+\sinh \left(\frac{\sqrt{\omega_{0} }}{\sqrt{2}}\right)}{\cos \left(\frac{\sqrt{\omega_{0} }}{\sqrt{2}}\right)+\cosh \left(\frac{\sqrt{\omega_{0} }}{\sqrt{2}}\right)} \right)+\mathcal{O} (\mathrm{Wo}^{4}),
\end{aligned}
\end{equation}
which is the same as formula \eqref{eq:homogenization effdiffusivity Shear1} obtained by the homogenization approach. We also can compute the asymptotic expansion in the high  Womersley number limit $\mathrm{Wo} \rightarrow \infty $ which yields
\begin{equation}
\begin{aligned}
  \tilde{\kappa}_{\mathrm{eff}}=&1+\frac{\mathrm{Pe}^2 \mathrm{Wo}^{2}}{2 \sqrt{2}} \left\{\frac{\sinh \left(\sqrt{2\omega_0 }\right)-\sin \left(\sqrt{2\omega_0 }\right)}{\sqrt{\omega_0 } \left(\omega_0 ^2-\text{Wo}^4\right) \left(\cos \left(\sqrt{2\omega_0 }\right)-\cosh \left(\sqrt{2\omega_0 }\right)\right)}\right.\\
  & \hspace{2cm}\left. -\frac{1}{\text{Wo}^5-\text{Wo} \omega_0 ^2}   \right\}+ \mathcal{O}\left( e^{-\frac{\sqrt{2}}{2}  \mathrm{Wo}} \right).
\end{aligned}
\end{equation}
Either low viscosity or large gap thickness yields a large Womersley number. In the low viscosity limit, since no fluid motion is generated for a parallel wall moving in an ideal fluid, the boosted diffusivity vanishes.
The numerical simulation results in figure \ref{fig:DifferentNu} show that the mixing is confined in a thinner boundary layer for a smaller viscosity.

Next, we study the limiting cases involving the non-dimensional frequency $\omega_0$ with fixed Womersley number. In other words, we change $\omega_0$ while keeping the spatial shape of the Stokes layer unchanged. As $\omega_0\rightarrow 0$, we have
\begin{equation}\label{eq:ekappaStokesLowFrequency}
\begin{aligned}
\tilde{\kappa}_{\mathrm{eff}}= &1+\frac{\text{Pe}^2}{2 \text{Wo}^2 \left(\cosh \left(\sqrt{2} \text{Wo}\right)-\cos \left(\sqrt{2} \text{Wo}\right)\right)}\left(-\frac{\sin \left(\sqrt{2} \text{Wo}\right)+\sinh \left(\sqrt{2} \text{Wo}\right)}{\sqrt{2} \text{Wo}}\right. \\
  &\left.+ \frac{\cos \left(\sqrt{2} \text{Wo}\right)+\cosh \left(\sqrt{2} \text{Wo}\right)+2 \cos \left(\frac{\text{Wo}}{\sqrt{2}}\right) \cosh \left(\frac{\text{Wo}}{\sqrt{2}}\right)+2}{3}
\right)+\mathcal{O} (\omega_{0}^{2}). \\
\end{aligned}
\end{equation}
We have the following asymptotic expansion as $\omega_{0}\rightarrow \infty$:
\begin{equation}\label{eq:ekappaStokesHighFrequency Fixed Wo}
\begin{aligned}
\tilde{\kappa}_{\mathrm{eff}} =&1+ \frac{  \mathrm{Pe}^2 \mathrm{Wo}^2}{2 \sqrt{2} \left( \cosh (\sqrt{2} \mathrm{Wo})-\cos (\sqrt{2} \mathrm{Wo}) \right)}
 \left\{
\frac{\sin (\sqrt{2} \mathrm{Wo})+\sinh (\sqrt{2} \mathrm{Wo})}{    \mathrm{Wo} \omega_{0} ^2} 
\right.\\
&\left.- \left( \cos (\sqrt{2} \mathrm{Wo})+\cosh (\sqrt{2} \mathrm{Wo})+2 \right)\omega_{0}^{-5/2}+\mathcal{O}\left( \omega_{0}^{-7/2} \right) \right\}.\\
\end{aligned}
\end{equation}

One may be interested in $\tilde{\kappa}_{\mathrm{eff}}$ as $\omega_0\rightarrow \infty$ or $\omega_{0}\rightarrow 0$ for a given Schmidt number $\mathrm{Sc}$. In this case, the Stokes shear wave becomes a steady flow $u (t,y)=y +\mathcal{O} (\omega_0)$ as $\omega_{0}\rightarrow 0$. Equation \eqref{eq:ekappaStokes} becomes the classical result of Taylor dispersion for a steady moving wall
\begin{equation}
\begin{aligned}
\tilde{\kappa}_{\mathrm{eff}}= &1+ \mathrm{Pe}^2 \left( \frac{1}{240}+\frac{\omega_{0} ^2 \left(7-155 \text{Sc}^2\right)}{3628800 \text{Sc}^2} \right) +\mathcal{O}\left(\omega_{0} ^{5/2}\right). \\
\end{aligned}
\end{equation} 
When $\omega_0\rightarrow \infty$, we have
\begin{equation}\label{eq:ekappaStokesHighSc}
\begin{aligned}
\tilde{\kappa}_{\mathrm{eff}}=  &1+\frac{\text{Pe}^2 \text{Sc}}{2 \sqrt{2} \left(\sqrt{\text{Sc}}+1\right) (\text{Sc}+1) \omega_{0} ^{3/2}} +\mathcal{O} \left( e^{-\min (1, \frac{1}{\sqrt{\mathrm{Sc}}}) \sqrt{2\omega_{0}}} \right).\\
\end{aligned}
\end{equation}

\begin{figure}
  \centering
    \includegraphics[width=1\linewidth]{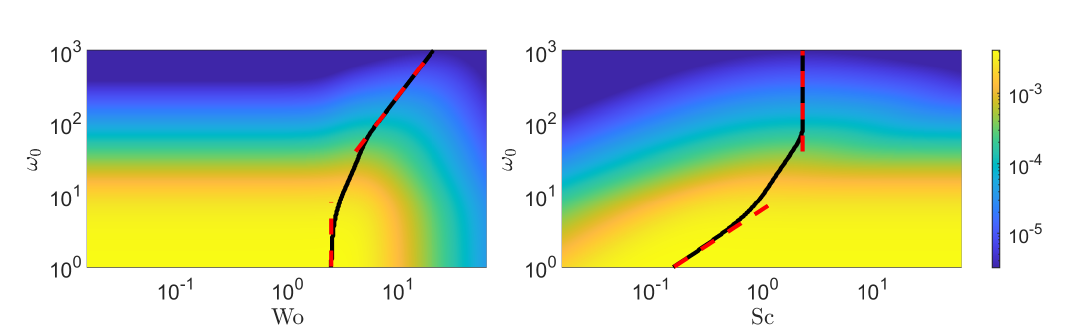}
  \caption{ Enhanced diffusivity  $\tilde{\kappa}_{\mathrm{eff}}-1$ for P\'{e}clet number $\mathrm{Pe}=1$, (left panel) varying the dimensionless frequency $\omega_{0}$ and the Womersley number $\mathrm{Wo}$ or (right panel) varying the dimensionless frequency $\omega_0$ and the Schmidt number $\mathrm{Sc}$. The black curves indicate the location of the enhanced diffusivity maximum in the non-dimensional parameter space(s) for a given non-dimensional frequency.  The red dashed curves are the asymptotic approximation of these functions for large or small $\omega_{0}$.  }
  \label{fig:EnhancedKappaVsWoSc}
\end{figure}

These asymptotic expansions imply the potential existence of a maximum effective diffusivity as Schmidt number or Womersley number is varied when $\omega_0$ is given. We denote the Schmidt number and Womersley number for reaching the maximum of $\tilde{\kappa}_{\mathrm{eff}}$ as $f_{\mathrm{Sc}} (\omega_{0})$ and $f_{\mathrm{Wo}} (\omega_{0})$ respectively. When $\omega_0$ is large, equation \eqref{eq:ekappaStokesHighSc} leads to
\begin{equation}\label{eq:ekappaStokesHighFoptimalSC}
\begin{aligned}
f_{\mathrm{Sc}} (\omega_0)\sim & \frac{1}{3}\left(\sqrt[3]{53+6 \sqrt{78}}+\sqrt[3]{53-6 \sqrt{78}}+2\right)\approx 2.3146, \; \omega_0\rightarrow \infty.\\
\end{aligned}
\end{equation}
When $\omega_0$ is small, we numerically calculate the maximum using \eqref{eq:ekappaStokesLowFrequency} and find
\begin{equation}\label{eq:ekappaStokesLowFoptimalWo}
\begin{aligned}
f_{\mathrm{Wo}} (\omega_0)\sim & 2.49426, \;  \omega_0 \rightarrow 0.\\
\end{aligned}
\end{equation}
The results of other cases can be obtained by the relation $\omega_0=\mathrm{Wo}^{2}  \mathrm{Sc}$. Figure \ref{fig:EnhancedKappaVsWoSc} shows how the enhanced diffusivity varies for different dimensionless parameters. The black curves represent the functions $f_{\mathrm{Wo}} (\omega_{0}), f_{\mathrm{Sc}} (\omega_{0})$ and the red dashed curve represents their asymptotic results.   

\begin{figure}
  \centering
 \includegraphics[width=.5\linewidth]{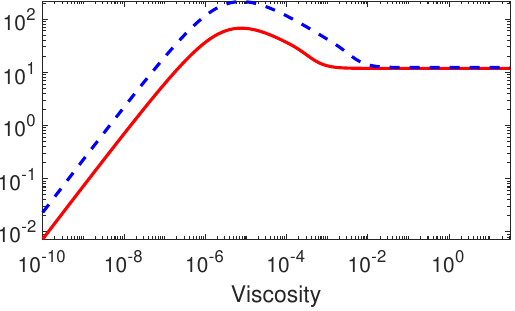}
  \caption[]
  {The dimensionless enhanced diffusivity  $\kappa_{\mathrm{eff}}-1$ versus the viscosity with parameters  $L=0.2$ cm, $A=1$ cm, $\kappa=3.3*10^{-6}$ $\text{cm}^2/s$,  $\omega=2\pi/100$ $rad/s$  (red solid curve, $\mathrm{Pe}=3808$), $\omega=2\pi/10$ s$^{-1}$ (blue dashed curve, $\mathrm{Pe}=38080$)}
  \label{fig:RelativeEnhancedDiffusivityVsViscosity}
\end{figure}

To further explore the maximal properties, we plot in figure \ref{fig:RelativeEnhancedDiffusivityVsViscosity} the normalized enhanced diffusivity as a function of the fluid kinematic viscosity with experimental parameters.  As the viscosity increases, the effective diffusivity first reaches its maximum value then decreases to a plateau. The difference between the peak and the plateau is smaller for smaller frequencies. Due to this phenomenon, it is hard to distinguish the maximum and plateau value of $\tilde{\kappa}_{\mathrm{eff}}$ at small frequencies in figure \ref{fig:EnhancedKappaVsWoSc}.

\begin{figure}
  \centering
 \includegraphics[width=.5\linewidth]{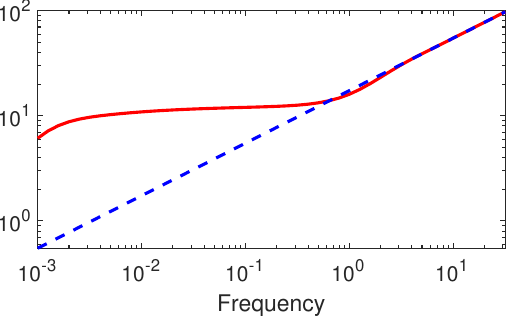}
  \caption[]
  {Comparison of dimensionless enhanced diffusivity ${\kappa_{\mathrm{eff}}}-1$ computed by the full expression of $\kappa_{\mathrm{eff}}$ in equation \eqref{eq:ekappaStokes} (solid red)  with the one computed by the high frequency asymptotic expansion of $\kappa_{\mathrm{eff}}$ given in equation \eqref{eq:ekappaStokesHighFrequency} (dashed blue), for the Stokes layer solution with parameters  $L=0.2$ cm, $A=1$ cm, $\kappa=3.3*10^{-6}$ $\text{cm}^2/s$, $\nu=0.01$ St ($\mathrm{Sc}=3030.3$). }
  \label{fig:DiffusivityVsFrequency}
\end{figure}

All of these results are obtained with a fixed \textrm{Pe}, which occurs as the amplitude $A\rightarrow 0$ as $\omega\rightarrow \infty$. Hence, in those cases, the effective diffusivity vanishes for large frequency. Things are different in dimensional variables.  Figure \ref{fig:RelativeEnhancedDiffusivityVsViscosity} suggests that higher dimensional frequency may yield higher effective diffusivity for a fixed amplitude $A$. Based on this observation, we are next interested in studying the effective diffusivity at large frequencies while holding all other physical parameters constant. For linear shear flow, the dimensional effective diffusivity $\kappa _{\mathrm{eff}}$ is bounded by a constant set solely by the gap thickness $L$ and the amplitude of wall motion $A$, 
\begin{equation}\label{eq:diffusivityShearUpperBound}
\begin{aligned}
\kappa_{\mathrm{eff}}\leq \kappa \left( 1+ \frac{\mathrm{Pe}^2}{2 \omega_{0} ^2} \right) = \kappa \left( 1+\frac{A^{2}}{2L^{2}} \right),
\end{aligned}
\end{equation}
 which follows from equation \eqref{eq:effdiffusivityShear}.  Alternatively, at finite viscosities, the Stokes wave solution induces an effective diffusivity which is unbounded in the high frequency limit $\omega \rightarrow \infty$ and has the following asymptotic expansion:
\begin{equation}\label{eq:ekappaStokesHighFrequency}
\begin{aligned}
   \kappa_{\mathrm{eff}} = \kappa \left( 1+ \frac{A^2 \nu  \sqrt{\omega }}{2 \sqrt{2} L \left(\sqrt{\kappa }+\sqrt{\nu }\right) (\kappa +\nu )} \right)+\mathcal{O} \left( e^{-\min (\frac{1}{\sqrt{\kappa}}, \frac{1}{\sqrt{\nu }}) L\sqrt{2\omega}} \right).
\end{aligned}
\end{equation}
The log-log plot (\ref{fig:DiffusivityVsFrequency}) shows the exponential convergence of $\kappa _{\mathrm{eff}}$ to its high frequency asymptotic expansion \eqref{eq:ekappaStokesHighFrequency}. 
One may be interested in whether such growth of the variance as a function of high frequency is visible at large but finite times.  This is a question which involves commuting limits and joint asymptotic expansion.  Careful examination of the formula in equation \eqref{eq:SecondMomentStokes} shows the high frequency expansion at fixed time produces a linearly growing term in time whose slope exactly matches that in equation (\ref{eq:ekappaStokes}) as well as the correction which is bounded in both frequency and time.  Hence, the time and high-frequency limits will commute in this case. There could be cases of incommensurate limits amongst the non-dimensional parameters.

\begin{figure}
  \centering
\subfigure[Schmidt number]{    \includegraphics[width=0.40\linewidth]{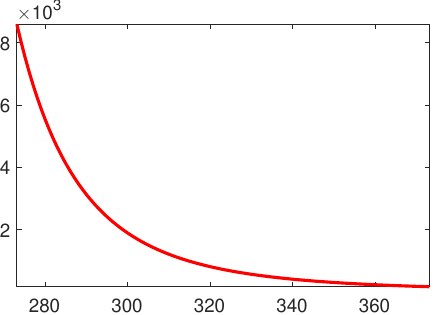}}
  \subfigure[$\kappa_{\mathrm{eff}} (\theta), \tilde{\kappa}_{\mathrm{eff}} (\theta)$]{    \includegraphics[width=0.46\linewidth]{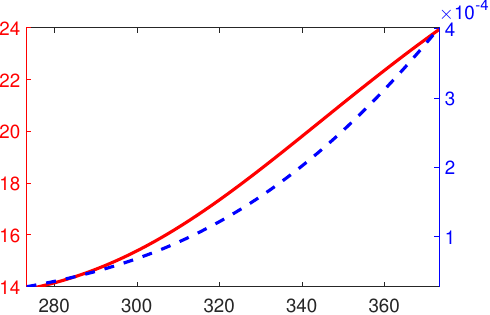}}
  \caption{ Panel (a) Schmidt number of fluorescein-water system varies with the temperature $\theta \in [273, 373] K$. Panel (b) $\tilde{\kappa}_{\mathrm{eff}} (\theta)$ (left $y$ axis, red color), $\kappa_{\mathrm{eff}} (\theta)$ (right $y$ axis, blue color) with parameters $A=1$ cm, $L=1/5$ cm, $\omega=2\pi/10$ rad$/s$.  }
  \label{fig:EnhanceDiffusionTemperature}
\end{figure}

The fluid viscosity and tracer diffusivity are both functions of temperature. For instance, they may satisfy the Stokes-Einstein relationship (page 320 of the book \cite{dill2012molecular}) $\kappa (\theta)= \frac{k\theta}{6\pi \eta (\theta) r}$, where $k= 1.3807\times 10^{-23} J\cdot K^{-1}$ is the Boltzmann constant, $r$ is the hydrodynamic radius of the tracer, $\eta$ is the dynamic viscosity, and $\theta$ is the absolute temperature with the unit Kelvin  $K$. Of course, this relationship is correct for a small spherical particle experiencing Brownian motion: the solute is a molecule, and not a sphere.  Still, measuring the diffusivity at one temperature can be nonetheless used to calculate an effective hydrodynamic radius.  Hence, equation \eqref{eq:ekappaStokesHighFoptimalSC} and \eqref{eq:ekappaStokesLowFoptimalWo} could provide good guidance for finding the temperature for the maximum of $\tilde{\kappa}_{\mathrm{eff}} (\theta)$. Since $\kappa_{\mathrm{eff}} (\theta) =\tilde{\kappa}_{\mathrm{eff}} (\theta)\kappa (\theta)$, we should also notice that the temperature for reaching the maximum of $\kappa_{\mathrm{eff}} (\theta)$ and $\tilde{\kappa}_{\mathrm{eff}} (\theta)$ could be different. We consider the case of the fluorescein diffusion in water. As a function of the temperature, the diffusivity of fluorescein  takes the form  $\kappa (\theta)=\frac{1.2717*10^{-8} \theta }{e^{\frac{578.919}{\theta -137.546}-3.7188}}$ cm$^{2}$/s \cite{rhodamineabsolute}, the dynamic viscosity of water is $\eta (\theta)=2.4152\times 10^{-4}\times 4.7428^{\frac{365.33}{-139.86 + \theta}}$ Poise (Table 2 in \cite{fogel2001temperature}), and the density of water \cite{kell1975density,jones1992its} is
\begin{equation}
\begin{aligned}
  \rho (\theta)= & \frac{10^{-3}}{0.0168979 (\theta -273)+1}( 999.84+16.9452 (\theta -273)\\
  &-0.00798704 (\theta -273)^2 -0.0000461705 (\theta -273)^3  \\
  &\left. + 1.0556302\times 10^{-7} (\theta -273)^4
-2.8054253\times 10^{-10} (\theta -273)^5\right) \text{ g/cm}^{3}.
\end{aligned}
\end{equation} With these formulas, we plot the Schmidt number as a function of temperature for $\theta \in [273,373]$ K in the panel (a) of figure \ref{fig:EnhanceDiffusionTemperature}.  To observe an interior maximum in the effective diffusivity, the Schmidt number must be smaller than $2.3146$.  For fluorescein-water mixtures, the minimum Schmidt number over this range of temperatures is $172.2862$, and thus, no interior maximum is observed. In fact, over this range of temperatures, $\tilde{\kappa}_{\mathrm{eff}} (\theta)$ increases monotonically as seen in panel (b) of figure \ref{fig:EnhanceDiffusionTemperature}. A tracer-fluid system with a Schmidt number smaller than $2.3146$ could exhibit an interior effective diffusivity maximum as a function of temperature.

\subsection{ Skewness}
In this section, we utilize the formulae derived in section \ref{sec:FrameDiffusivitySkewness} to study the skewness of $\bar{T}$ for left-right symmetric initial data.

At infinite viscosity, the Stokes layer solution \eqref{eq:StokeswaveVelocityNon} becomes a periodic time-varying linear shear flow $y\xi (t)$.  It is fairly straightforward to show that the passive scalar skewness is generally zero for initial data $\delta (x)$ by the analysis of parity.  Observe that the linear shear admits an odd cosine expansion in $y$ and produces an odd $T_1$ cosine expansion in $y$.  In turn, we see that $T_2$ is even from inspection, since the driver in the equation for $T_2$ is the product of two functions $u$ and $T_1$ which are odd about the centerline of the channel $y = 1/2$.  Lastly, the driver for the $T_3$ equation contains $T_1$ (odd) and the product of $\tilde{u}$ (odd) and $T_2$ (even).  When computing the net third moment by cross-sectional averaging, $\bar{T}_1=0$ as well as $ \overline{\tilde{u}T_2} =0$.  Hence, the skewness is zero for a linear shear.  Alternatively, it is easy to check that $\left\langle (y- \frac{1}{2})\phi_{n_{1}},\phi_{n_{2}} \right\rangle \left\langle (y- \frac{1}{2}), \phi_{n_{2}} \right\rangle=0$ for any pair of $\left( n_{1},n_{2} \right)$. Then, we also see the skewness is zero for all time from equation \eqref{eq:ThirdArisMoment}.

\begin{figure}
  \centering
    \includegraphics[width=0.5\linewidth]{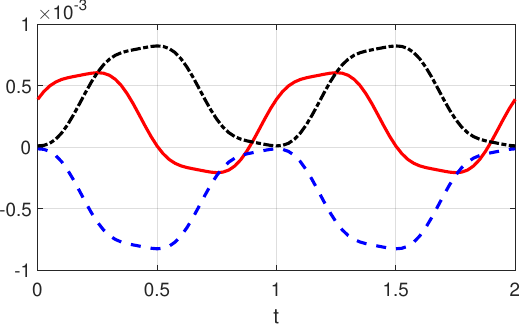}
  \caption{The coefficient of $t^{-{3}/{2}}$ in the long-time asymptotic expansion of the skewness of $\bar{T}$ (i.e. equation \eqref{eq:skewnessSingleF}) with the parameters $\mathrm{Pe}=2, \mathrm{Wo}=1, \omega_0= 2 \pi$ and the velocity of the wall $\cos (\omega_{0} t+s)$. The red solid curve, blue dash curve and black dash-dot curve correspond to the phase shift $s=0, s= {\pi}/{2}, s=-{\pi}/{2}$, respectively.   }
  \label{fig:Skewness}
\end{figure}

At finite viscosities, the skewness of $\bar{T}$ has more interesting behavior. With the formula for the Stokes layer solution \eqref{eq:StokeswaveVelocityNon}, we have 
\begin{equation}
\begin{aligned}
&\left\langle u_{k_{1}},\phi_{n_{1}} \right\rangle=\tfrac{e^{\mathrm{i} \frac{\pi}{4}} \sqrt{k_{1}} \mathrm{Wo}  \left((-1)^{n_{1}} \cosh \left(e^{\mathrm{i} \frac{\pi}{4}} \sqrt{k_1} \mathrm{Wo}\right)-1\right)}
  {\sqrt{2}\mathrm{sinh}\left(e^{\mathrm{i} \frac{\pi}{4}} \sqrt{k_1} \mathrm{Wo}\right)\left( \pi ^2 n_{1}^2+\mathrm{i} k_1 \mathrm{Wo}^2 \right)}, \\
 &\left\langle u_{k_{3}}, \phi_{n_{2}}  \right\rangle =\tfrac{e^{\mathrm{i} \frac{\pi}{4}} \sqrt{k_{3}} \mathrm{Wo}  \left((-1)^{n_{2}} \cosh \left(e^{\mathrm{i} \frac{\pi}{4}} \sqrt{k_3} \mathrm{Wo}\right)-1\right)}
  {\sqrt{2}\mathrm{sinh}\left(e^{\mathrm{i} \frac{\pi}{4}} \sqrt{k_3} \mathrm{Wo}\right)\left( \pi ^2 n_{2}^2+\mathrm{i} k_3 \mathrm{Wo}^2 \right)}, 
\\
 &\left\langle u_{k_2}\phi_{n_{1}},\phi_{n_{2}} \right\rangle=\tfrac{e^{\frac{\mathrm{i} \pi }{4}} \sqrt{k_2} \text{Wo} \left(k_2 \text{Wo}^2-\mathrm{i} \pi ^2 \left(n_1^2+n_2^2\right)\right) \left(1+(-1)^{1+n_{1}+n_2} \cosh \left(e^{\frac{\mathrm{i} \pi }{4}} \sqrt{k_2} \text{Wo}\right)\right)}
{\left(-2 \pi ^2 k_2 \left(n_1^2+n_2^2\right) \text{Wo}^2-\mathrm{i} k_2^2 \text{Wo}^4+\mathrm{i} \pi ^4 \left(n_1^2-n_2^2\right){}^2\right) \sinh \left(e^{\frac{\mathrm{i} \pi }{4}} \sqrt{k_2} \text{Wo}\right)}.\\
\end{aligned}
\end{equation}
Therefore the formula of $S (\bar{T})$ is available by applying formula \eqref{eq:ThirdArisMoment} and \eqref{eq:SkewnessDefinition}.  Figure \ref{fig:Skewness} shows the coefficient of $t^{-\frac{3}{2}}$ in the long time asymptotic expansion of $S (\bar{T})$ with the wall velocity $\cos (2\pi t+s)$.  As predicted by equation \eqref{eq:skewnessSingleF}, the sign of the skewness changes periodically. The skewness sign stays positive longer than   negative when the  phase shift $s$ of the wall motion  is zero. However, it stays strictly positive when $s=\pi/2$ and strictly negative when $s=-\pi/2$. This observation suggests that we can control the symmetry properties of $\bar{T}$ by simply shifting the phase. In addition, figure \ref{fig:Skewness} shows that the skewness is not zero at the end of each period when the wall goes back to the initial position. The  numerical simulation results in figure \ref{fig:DifferentNu} also show that the distribution of tracer is asymmetric about the centerline of the initial data $x=0$. This phenomenon implies that, even with periodic flow in time, the symmetry of the tracer's distribution may break in the presence of diffusion.  Also note that, upon close inspection of the linear shear case documented in figure \ref{fig:StokesDiffernetFrame} and \ref{fig:CompareDiffusionEndTime} , one can see broken symmetry near the top and bottom of the graphs, though when cross-sectionally averaged, this effect cancels.

We also are interested in the short time behavior of the skewness. Article \cite{aminian2015squaring} presented a method for computing the short-time asymptotics of the Aris moment in an arbitrary cross-sectional domain. They found there is a plateau of skewness of $\bar{T}$ at short time which only depends on the geometry of the cross section.  They denoted this quantity as the geometric skewness. The geometric skewness is independent of the P\'eclet number. Hence, it can be computed by  neglecting the molecular diffusion.  For given initial data $T_{I}(x,y)$, the solution can be obtained by method of characteristics as $T(x,y,t)=T_{I}(x-\int\limits_0^t u(y,s)\mathrm{d}s, y) $, then $ \bar{T}_n = \int\limits_{-\infty}^{\infty}x^n\int\limits_{0}^{1} T_{I}(x-\int\limits_0^t u(y,s)\mathrm{d}s, y)\mathrm{d}y \mathrm{d}x $.  For general initial data, this leads to a lengthy analytical formula for the geometric skewness, which is too long to list here.  We will study its behavior in section \ref{sec:ExperimentAndTheoreticalResults} and compare with computational simulations, which will also show that the skewness depends significantly on the phase shift at short times. 


\section{Computational approaches}
In this section, we describe two computational approaches for solving the advection-diffusion equation: the  Monte-Carlo method and the Fourier spectral method. The Monte-Carlo method is advantageous to problems involving complex geometry and is ideally suited to parallel computing. Moreover, its convergence rate only depends on the number of samples which makes it particularly useful for higher-dimensional integrals.  Based on those features, the Monte-Carlo methods are more suitable for computing the Aris moments on larger time scales. However, it is expensive to store the positions of millions of particles at every observation time instant.  The spectral method is more efficient and flexible to compute the distribution of the tracer for different parameters on a shorter time scale, which can remedy the weakness of Monte-Carlo method.

 First, we introduce the setup of the  Monte-Carlo method.  The Monte-Carlo simulations are used to compare with the laboratory experiments described in the following section. To get a global approximation of the solution of the advection-diffusion equation, we adopt the forward Monte-Carlo method which is based on the Fokker-Planck equation.  We determine the initial position of $10^7$ particles according to the intensity distribution of the experimental photographs on a uniform grid.  We assume that the tracer is uniformly distributed on the cross section of the channel.  Each particle's trajectory satisfies the stochastic differential equation (SDE),
\begin{equation}
\begin{aligned}
&\mathrm{d}X_{t}=u(Y_{t},t)\mathrm{d}t+ \sqrt{2 \kappa} \mathrm{d}W_{1},\quad \mathrm{d}Y_{t}=\sqrt{2 \kappa}\mathrm{d}W_{2},\quad \mathrm{d}Z_{t}=\sqrt{2 \kappa}\mathrm{d}W_{3}.
\end{aligned}
\end{equation}
where $u(y,t)$ is a shear flow, $\kappa$ is the molecular diffusivity and $\mathrm{d}W_i$ are independent white noises. We solve the SDE by the Euler scheme with a time increment $\Delta t =0.05$ s which resolves the frequencies studied experimentally,
\begin{equation}
\begin{aligned}
&X_{t_{i+1}}=X_{t_{i}}+u(Y_{t_{i}},t_{i})\Delta t+ \sqrt{2 \kappa \Delta t}n_{i,1},\\
&Y_{t_{i+1}}=Y_{t_{i}}+\sqrt{2 \kappa \Delta t}n_{i,2},  \\
&Z_{t_{i+1}}=Z_{t_{i}}+\sqrt{2 \kappa \Delta t}n_{i,3}.\\
\end{aligned}
\end{equation}
Here, $n_{i,j}$ are independent and identically distributed standard normal random variables which are produced by the Mersenne Twister uniform random number generator and Marsaglia polar method \cite{marsaglia1964convenient}. We impose the billiard-like reflection rules on the boundary plane $z=0$ cm, $z=16$ cm, $y=0$ cm, $y=L$. We note that the tank height is chosen to be $16$ cm to match the experimental height.  At a given time $t$, the histogram of the $N=10^{7}$ particle positions is an approximation of the solution $T(x,y,z,t)$. The cross-sectional average of $n$th Aris moment can be approximated by the formula
\begin{equation}
\begin{aligned}
 \bar{T}_n (t_{i})  =& \frac{1}{N} \sum\limits_{j=1}^N X_{t_{i},j}^n, \\
\end{aligned}
\end{equation}
where $ X_{t_{i},j}$ is the $x$-coordinate of $j$th particle at time $t_{i}$. The simulations are performed on UNC's Longleaf computing cluster by using 200 cores.  The computation takes approximately $8$ h to perform $3\times10^5$ time steps needed to resolve the flow and reach the diffusion timescale $L^2/ \kappa$.

	Additionally, we utilize the Fourier spectral method to solve the two-dimensional advection-diffusion equation \eqref{eq:Advection Diffusion Equation deterministic shear} with Stokes layer solution \eqref{eq:StokeswaveVelocityNon}. All computations of solution and Aris moments are performed on the domain $[-H,H]\times[0,L]$. When $H$ is large enough, we can assume there is a periodic boundary condition in the $x$-direction. Since there are non-penetration conditions in the $y$-direction, we perform the even extension in the $y$-direction to obtain the periodic condition on the extended domain.  Thus, we solve the advection-diffusion equation with periodic boundary conditions on the rectangular domain $[-H, H]\times[0, 2L]$. It can be solved by the standard Fourier spectral method with the explicit fourth order Runge-Kutta method as the time-marching scheme. In the dealiasing process at each time step, we apply the all-or-nothing filter with the two-thirds rule to the spectrum; that is, we set the upper one-third of the resolved spectrum to zero (see chapter 11 of the book \cite{boyd2001chebyshev} for details). We solve equation with the parameters $H=16$ cm, $L=0.2$ cm, and time increment $\Delta t= 0.005$ s over $2000$ time steps. The grid resolution is $2048 \times 257$ before the even extension and  $2048 \times 512$ after the extension.

\section{Experimental methods}

\subsection{Experimental setup}

Experiments were performed in a $50 \times 25 \times 30$ cm glass tank. To reduce effects of thermal convection, the fluid was density stratified using the two bucket method
\cite{economidou2009density,oster1965density}
with sodium chloride as the stratifying agent. The density of the background fluid linearly decreases with height, with total variation approximately $0.1$ g/cc over $20$ cm. One wall, made of $0.75$ in thick glass, is fixed to both sides of the tank, while a second $0.25$ in thick aluminum wall is connected from above to a linear stage driven by an Oriental motor model ARM66MC with driver model ARD-A, which translates the wall in the horizontal direction parallel to the fixed wall. The motor is controlled by custom software written in MATLAB for the ATMEL ATMEGA2560 microcontroller and implemented using an Arduino MEGA 2560. To prepare the tracer, fluorescein powder is mixed with saline solution of density $1.05$ g/cc to a concentration of $0.9$ g/L. About $50$ $\mu$L of fluorescein solution is injected between the walls near the center of the interrogation region and allowed to freely diffuse for several hours to make the dye uniformly in the cross-section. The tank and motor frame are draped in black fabric to block ambient light, and a blacklight is placed on top of the tank to illuminate the tracer. The illuminated fluorescein dye is photographed from the side using a Nikon D750 which is synchronized with the oscillating wall period using the Arduino. A first-surface mirror tilted back $45$ degrees from vertical is placed below the tank to allow for easily viewing the dye from below.

To capture particle tracking velocimetry (PTV) images, saline solution of density $1.05$ g/cc is mixed with $50$ micron diameter hollow glass microspheres and injected into the interrogation region. A laser sheet with normal in the vertical direction illuminates the fluid which is viewed from below using $30$ fps video captured on a Nikon D750 equipped with a Nikon AF-S micro Nikkor $105$ mm lens. PTV processing is performed in MATLAB using PTVlab \cite{brevis2011integrating}.  Figure \ref{fig:schematic} shows a schematic of the experimental setup from three different views.

\subsection{Image Analysis}
To process the dye images, a Gaussian filter is applied, and then the intensity is integrated along the vertical direction. Then the full width at half maximum (FWHM) is measured as a function of time, first for the case of no wall movement to measure the bare diffusivity of sodium fluorescein in the saline solution, then after turning on the wall to measure the effective diffusivity.

\begin{figure}
  \centering
    \includegraphics[width=1\linewidth]{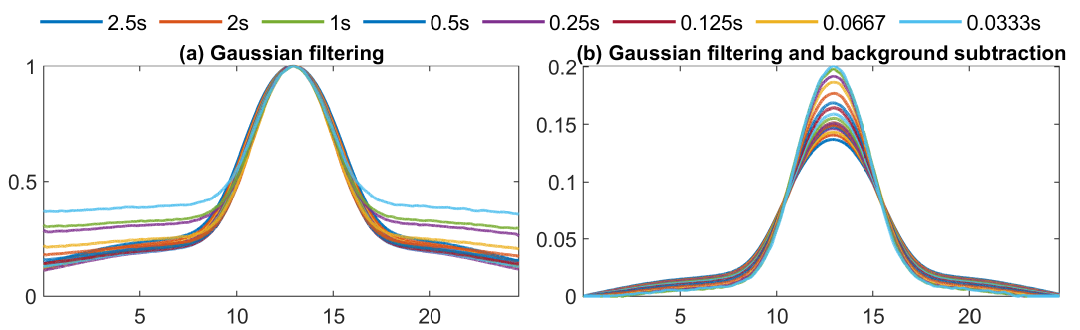}
  \caption{ \textbf{Study of the image noise:}  For a fixed experimental observation of dye concentration with no flow, we take photos at different shutter speeds and process the resulting $\bar{T}$, which effectively adjusts the signal to noise ratio while keeping the signal fixed.  Panel (a) We apply a 2-D Gaussian filter by the Matlab built-in function \texttt{imgaussfilt} with parameter \texttt{sigma}=25. Here each curve is rescaled to have maximum one. Panel (b) We apply the 2-D Gaussian filter with the same parameter and subtract the background noise from the images (subset of exposure times and associated curve colors indicated in top legend). Here each curve is normalized to be a PDF.	}
 \label{fig:StudyShutter}
\end{figure}

In a distribution, the FWHM statistic is the difference between the two values of the independent variable at which the dependent variable is equal to half of its maximum value. The motivation for using the FWHM statistic in lieu of moment based measurements is summarized in figure \ref{fig:StudyShutter} and table \ref{tab:StudyShutter}.  Photographs with different exposure times of the same dye distribution are taken after the dye has been diffusing for several hours.  This provides a sequence of images with different signal to noise ratios of the same dye concentration field.  Small noise in the far field gives a large contribution to the moments as we see a large variation of the variance computed by the moment integral method in the second row of table \ref{tab:StudyShutter}. To obtain a  measurement of variance that is more robust to noise, we can take advantage of the explicit formula of the tracer's distribution. 
The multiscale analysis in appendix \ref{sec: MultiscaleAnalysis} shows that $\bar{T}$ can be approximated as a normal distribution at long times,
\begin{equation}
\begin{aligned}
\bar{T}= & \frac {\bar{T}_{0}}{\sqrt{4\pi t\kappa_{\mathrm{eff}} }}  \exp \left( -\frac{(x-\bar{T}_{1})^{2}}{4t\kappa_{\mathrm{eff}} } \right).  
\end{aligned}
\end{equation}
Hence the relationship between FWHM and the effective diffusivity $\kappa_{\mathrm{eff}}$ is
\begin{equation}\label{eq:FWHMdefination}
\begin{aligned}
\mathrm  {FWHM}=2{\sqrt  {2\ln 2}} \sqrt{\bar{T}_{0}2t \kappa_{\mathrm{eff}}} \approx 2.355 \sqrt{\bar{T}_{0}2t \kappa_{\mathrm{eff}}}. \\
\end{aligned}
\end{equation}
The first row of table \ref{tab:StudyShutter} shows the FWHM is more robust to noise, particularly when the signal-noise ratio is small.  For these reasons, we adopt the FWHM for measuring the effective diffusivity.
\begin{table}
  \centering
  \begin{tabular}{c|cccc}
\hline
 Exposure Time (s)& 2nd Moment & 2nd Moment BNS & FWHM & FWHM BNS \\
\hline\hline
2.500& 27.562& 14.665& 7.774& 6.273 \\
2.000& 27.099& 14.123& 7.362& 5.989 \\
1.600& 26.607& 13.913& 6.999& 5.770 \\
1.300& 26.536& 13.378& 6.780& 5.563 \\
1.000& 26.362& 12.722& 6.582& 5.384 \\
0.769& 26.691& 13.503& 6.466& 5.287 \\
0.625& 26.138& 13.485& 6.247& 5.200 \\
0.500& 26.204& 13.915& 6.264& 5.207 \\
0.400& 26.336& 13.544& 6.264& 5.144 \\
0.333& 25.574& 13.178& 6.091& 5.050 \\
0.250& 25.148& 13.037& 6.117& 5.136 \\
0.200& 26.116& 12.039& 6.169& 4.995 \\
0.167& 25.888& 11.458& 6.023& 4.910 \\
0.125& 26.615& 10.457& 5.946& 4.788 \\
0.100& 27.626& 9.711& 5.845& 4.599 \\
0.077& 29.199& 8.527& 5.745& 4.306 \\
0.067& 31.456& 6.959& 5.887& 4.078 \\
0.050& 35.155& 6.397& 6.528& 3.877 \\
0.040& 36.377& 5.834& 6.626& 3.754 \\
0.033& 39.178& 6.063& 7.832& 3.561 \\
\hline
  \end{tabular}
\caption{The variance computed by various methods for the data presented in figure \ref{fig:StudyShutter}. Here, the label `BNS' indicates the background noise was subtracted. The label `FWHM' indicates the variance was calculated by the full width at half maximum method which is given in equation  \eqref{eq:FWHMdefination}, and `2nd Moment' indicates the variance was calculated by the second Aris full moment which is given in equation \eqref{eq:effectiveDiffusivityDefinition}.} \label{tab:StudyShutter}
\end{table}

 \section{Experimental and theoretical results}\label{sec:ExperimentAndTheoreticalResults}
\begin{figure}
  \centering
 \includegraphics[width=1\linewidth]{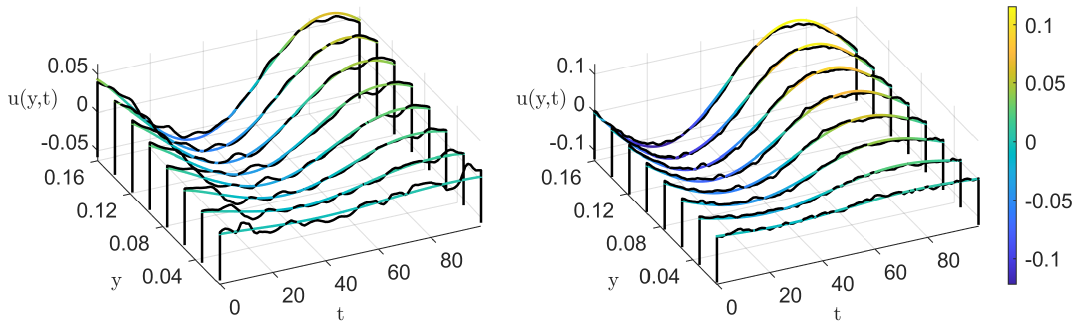}
  \hfill
  \caption[]
 {Comparison of particle tracking velocimetry (PTV) data (black curves) with the Stokes layer analytical solution (color curves) given in equation \eqref{eq:StokeswaveVelocityNon}. Each curve which is plotted by black curves corresponds with a time series of the shear velocity over a duration of one period taken at different distances between the fixed wall and the moving wall (located at L = 0.16 cm). Left panel has wall oscillation amplitude  $A = 1$ cm, right panel has  $A = 2$ cm, other parameters:  $\omega=2\pi /100$ rad$/s$, $\nu=0.0113$ St, and $L=0.16$ cm.}
  \label{fig:PTVmesh}
\end{figure}

Here, we present a comparison of experimental results with the theory developed above as well as Monte-Carlo and pseudo-spectral simulations for the evolving passive scalar field.  First, in figure \ref{fig:PTVmesh} we show an experimental and theoretical comparison of the Stokes layer \eqref{eq:2dStokesSol} for two different cases corresponding to two different amplitude wall motions.  The left panels show the shear velocity time series at $8$ different locations uniformly distributed across the channel for a case with $A=1$ cm, $\omega=2\pi \times0.01$rad/s, $\nu=0.0113$ St, and $L=0.16$ cm, while the right panels change the amplitude to $A=2$ cm. For the PTV experiment presented in figure \ref{fig:PTVmesh}, the Womersley number is $\mathrm{Wo}=0.16 \sqrt{\frac{2 \pi/100 }{0.0113}} \approx 0.3773$. The low Womersley number expansion in equation \eqref{eq:lowWoExpansion} would be a good approximation for the flow in this experiment.

\begin{figure}
  \centering
    \subfigure[Trial 3 in table \ref{tab:experimentResult}]{
       \includegraphics[width=1\linewidth]{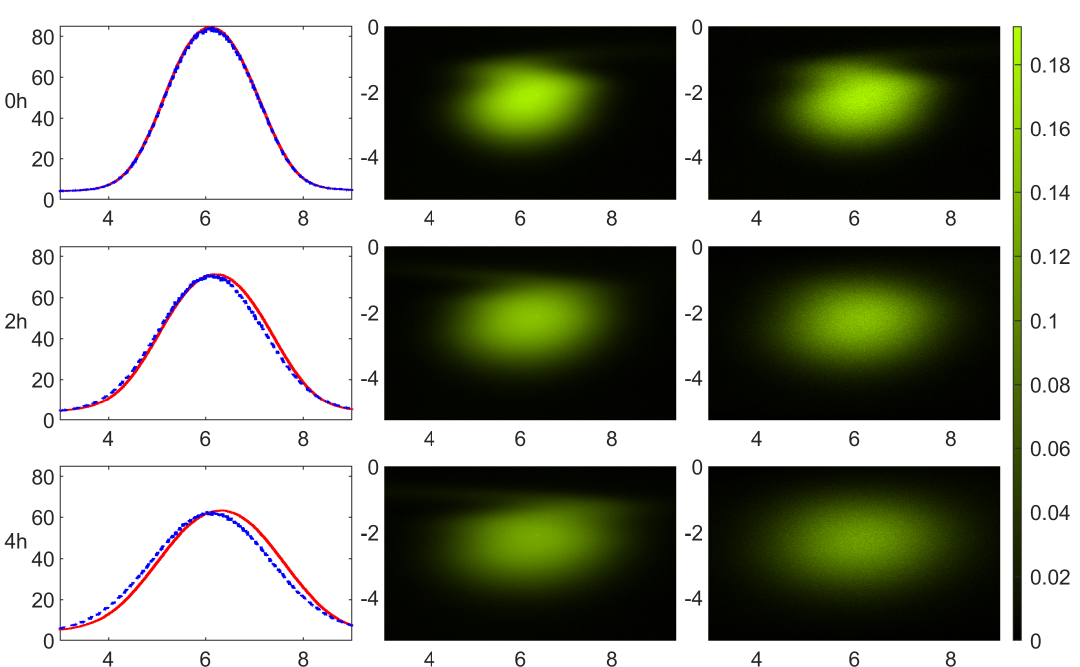}
    }
\subfigure[Trial 7 in table \ref{tab:experimentResult}]{
   \includegraphics[width=1\linewidth]{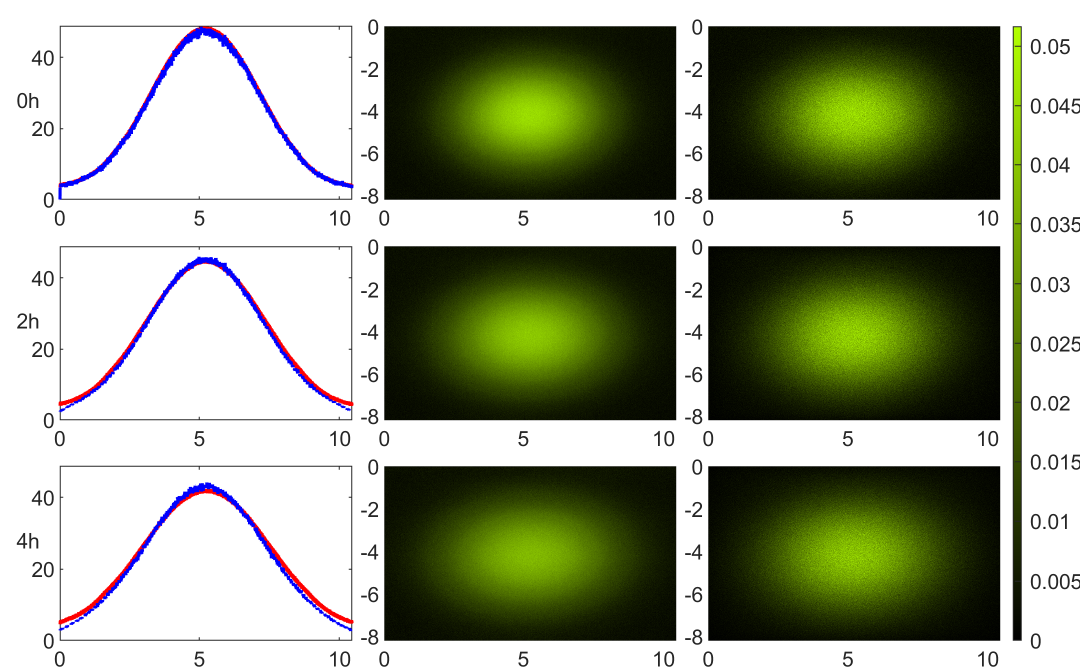}
}
\hfill
  \caption[]
{\textbf{Experimental and Monte-Carlo simulation comparison}: First column of panels shows the longitudinal distribution of the tracer $\bar{T}$, where the red solid line and blue dash line represent the experiment data and Monte-Carlo simulation, respectively. The second column of panels shows the experimental photographs of the tracer distributions viewed from the side at times $t=0, 2, 4$ h. The third column of the panels shows the corresponding Monte-Carlo simulations of the second column, where we also apply a 2-D Gaussian filter by the MATLAB built-in function \texttt{imgaussfilt} with parameter \texttt{sigma}=1. The parameters are $A = 1$ cm, $\nu=0.0113$ St, panel (a) $L=0.5$ cm, $\omega=2\pi/200$ $rad/s$, $\kappa=8.7\times10^{-6}$ cm$^{2}$/s, and panel (b) $L=0.35$ cm, $\omega=2\pi/400$ $\mathrm{rad}/s$, $\kappa=6.0\times10^{-6}$ cm$^{2}$/s.}
  \label{fig:3}
\end{figure}

Next, in figure \ref{fig:3} we show the experimental and Monte-Carlo simulations for the dye distribution viewed from the side at times $t=0$ s, $t=7,200$ s, and $t=14,400$ s, with parameters listed in the figure caption.  We also plot the averaged concentrations, $\bar T$ for the experiment and the simulation in the left columns of each panel.  The parameters for this figure correspond to trial $3$ (panel a) and $7$ (panel b) from table \ref{tab:experimentResult}.  A few comments regarding our experimental data.   First, since the width of the initial blobs is larger in panel b, the observed spreading is less than that in panel a even though the effective diffusivities are similar.  Second, in the absence of a flow, the cloud would have spread
at a much slower rate than those observed in this figure.
 
Table \ref{tab:experimentResult} shows the detailed comparison between the experimental campaign and theoretical prediction of the effective diffusivity.  First, we remark that the bare molecular diffusivity shows some variation.  This is primarily due to the unexpected dependence of fluorescein's diffusivity upon the concentration of NaCl which has been observed in other work by Gupta et al. \cite{gupta2019diffusion}.  In future work, we will explore this subtle effect.  Consequently, for the present case, we always measure the diffusivity first in our experiments.  

\begin{table}
  \centering
  \begin{tabular}{c|cccccccc}
\hline
Trial & $A$ & $\omega$ & $L$& $\rho$& $\kappa$&  $\kappa_{\mathrm{eff},e}$ &  $\kappa_{\mathrm{eff},t}$& Error\\
\hline\hline
1 &1 & $2\pi/200$& 0.3&$1.05$&   8.81E-06 &   4.26E-05&   5.39E-05 & 0.209 \\
2 &2&$2\pi/400$&0.3&$1.05$&8.81E-06 & 1.36E-04 &1.80E-05 &0.245\\
3&1 &$2\pi/200$&0.5&$1.05$&8.70E-06&2.71E-05&2.54E-05&-0.067\\
4&1& $2\pi/200$&0.3&$1.05$&8.26E-06&3.63E-05&5.06E-05&0.282\\
5&1&$2\pi/200$& 0.3&$1.03$&6.59E-06&2.97E-05&4.07E-05&0.270\\
6&1/5&$2\pi/400$& 0.35&$1.03$&6.75E-06&7.23E-06&7.76E-06&-0.068\\
7&1&$2\pi/400$& 0.35&$1.03$&5.985E-06&2.75E-05&2.84E-05&0.028\\
\hline
  \end{tabular}
\caption{  \textbf{Comparison of the experimental and theoretical effective diffusivity:} $A$(cm) is the amplitude of the wall motion, $\omega$ (rad$\cdot$ s$^{-1}$) is the frequency of the wall motion, $L$ (cm) is the gap thickness, $\rho$ (g/cc) is the local density, $\kappa$ (cm$^{2}$/s) is the molecular diffusivity measured from the pure diffusion stage in the experiment,  $\kappa_{\mathrm{eff},e}$ (cm$^{2}$/s) is the effective diffusivity computed by the FWHM approach from the experimental data, the viscosity is $\nu=0.0113$ St, and $\kappa_{\mathrm{eff},t}$(cm$^{2}$/s) is the theoretical value based on the experimental parameters. The last column is the relative error between experimental and theoretical effective diffusivity, $\frac{\kappa_{\mathrm{eff},t}-\kappa_{\mathrm{eff},e}}{\kappa_{\mathrm{eff},t}}$.} 
\label{tab:experimentResult}
\end{table}

\begin{figure}
  \centering
 \includegraphics[width=1.\linewidth]{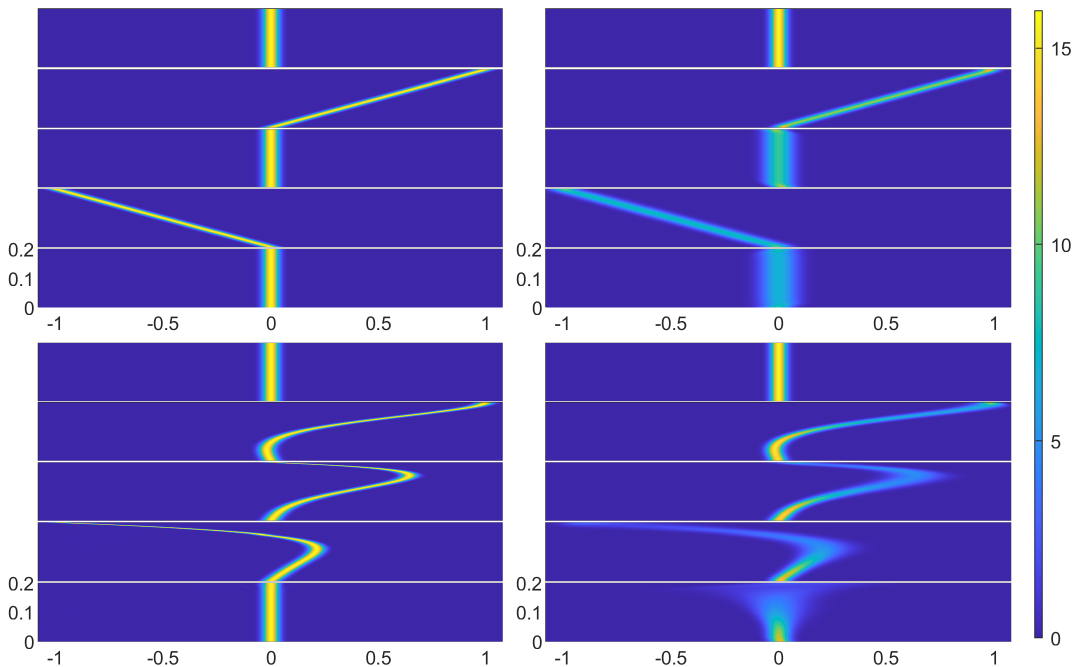}
  \caption[]
  {Spectral method comparison between mixing by linear shear versus nonlinear Stokes layer with a single-frequency sinusoidal wall motion.  Upper panels correspond to linear shear, while the lower panels correspond to the nonlinear Stokes layer, with parameters $\nu=0.001$ St, $\omega=0.2\pi$ rad$/s$, $L=0.2$ cm, $A=1$ cm.  The left panels are computed with $\kappa=0$ cm$^2$/s, while the right panels utilize $\kappa=10^{-5}$ cm$^2$/s.  Output times are taken at quarter periods, i.e., t=0 s, 2.5 s, 5 s, 7.5 s, 10s. }
  \label{fig:StokesDiffernetFrame}
\end{figure}

\begin{figure}
\centering
\subfigure[Top view of tracer]{
 \includegraphics[width=0.46\linewidth]{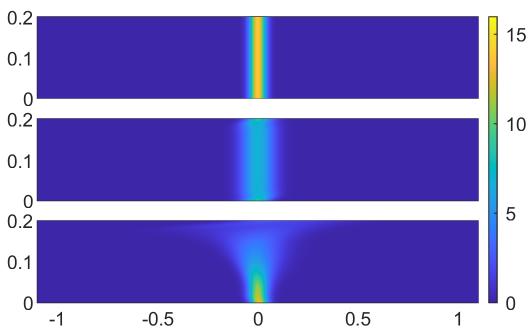}
}
\subfigure[$\int_{-\infty}^{\infty}\left| \partial_{y} T (x,y,10) \right| \mathrm{d}x$]{
 \includegraphics[width=0.46\linewidth]{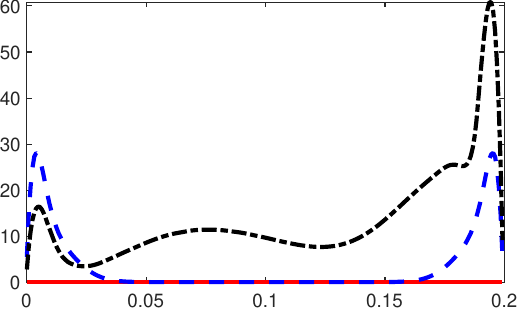}
}
  \caption[]
  {Spectral method comparison between mixing by different flows after one-period of motion  $t=10$ s.  Panel (a) The top panel has no flow, the middle has a linear shear, and the bottom panel has a nonlinear Stokes layer, with parameters $\nu=0.001$ St, $\omega=0.2\pi$ rad$/s$, $L=0.2$ cm, $A=1$ cm and  $\kappa=10^{-5}$ cm$^2$/s. Panel (b)Integral of the absolute value of the concentration gradient $\int_{-\infty}^{\infty}\left| \partial_{y} T (x,y,10) \right| \mathrm{d}x$, the red solid curve, blue dash curve, black dash dot curve correspond to no flow, linear shear flow, Stokes layer flow, respectively.  }
  \label{fig:CompareDiffusionEndTime}
\end{figure}

We can gain some insight into the transient effects giving rise to the long time limiting effective diffusion by studying the short time behavior using the spectral method with different diffusivities.  Shown in figure \ref{fig:StokesDiffernetFrame} are images of the scalar distributions, each case output at 5 different times taken on quarter cycles of the wall oscillation.  The top cases correspond to a pure time-varying linear shear with a single frequency sine wall motion, while the bottom panels correspond to cases with a nonlinear Stokes layer, with parameters $\nu=0.001$ St, $\omega=0.2\pi$ rad$/s$, $L=0.2$ cm, $A=1$ cm.  The left panels have zero diffusivity, while the right panels have $\kappa = 10^{-5}$ cm$^2$/s.  Observe in the case of the Stokes layer, the scalar is stretched into an extremely thin filament in the upper part of the channel which diffuses rapidly in the non-zero diffusivity case.  Compared to the linear shear, this case diffuses faster locally in the upper channel.  The case with linear shear is more uniformly mixed across the channel.  In the nonlinear Stokes layer case, the upper channel mixes very quickly.  This in turn increases the vertical concentration gradient, which gives rise to increased transient vertical diffusive tracer mixing.  To demonstrate this, we plot the integral of the absolute value of the vertical concentration gradient in the right panel of figure \ref{fig:CompareDiffusionEndTime}, for the cases examined in the left panel of zero, finite, and infinite viscosity showing that the finite viscosity Stokes layer has a significantly larger concentration gradient. This effect is perhaps more pronounced than in the more familiar steady pressure driven flow as a full cycle returns the Lagrangian map to its initial configuration.  

The left panel of figure \ref{fig:CompareDiffusionEndTime} shows the mixing result at the end of one period of wall motion for different flows. Flows create more dispersion in the longitudinal direction than the bare molecular diffusion. However, the physical mechanisms between a linear shear flow and a nonlinear Stokes layer flow give rise to very different enhanced diffusivities:  for the linear shear case, $\kappa_{\mathrm{eff}} =0.00013$ cm$^2$/s,  $13.14$ times the bare molecular diffusivity.  This value is nearly the upper bound \eqref{eq:diffusivityShearUpperBound} for a linear shear described above, which in this case is $13.5$ times the molecular diffusivity.  On the other hand, in the nonlinear Stokes layer case, $\kappa_{\mathrm{eff}}=0.00041$ cm$^2$/s, which is $40.96$ times the bare molecular diffusivity. 

\begin{figure}
  \centering
 \includegraphics[width=1.\linewidth]{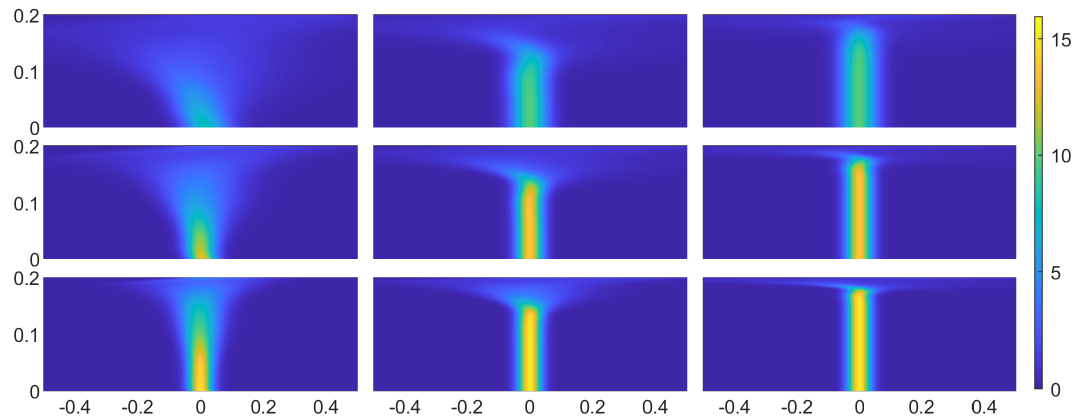}
  \caption[]
  {Spectral method comparison between mixing by Stokes layer flows after one-period of motion  $t=10$ s for  $\omega=0.2\pi$ rad$/s$, $L=0.2$ cm, $A=1$ cm, different diffusivities and viscosities.  The viscosity decreases from left to right ($\nu$ =0.01 St, 0.001 St, 0.0001 St) and the diffusivity decreases from the top to bottom ($\kappa=5\times10^{-5}$ $\text{cm}^{2}$/s, $10^{-5}$ cm$^{2}$/s, $2\times10^{-6}$ cm$^{2}$/s). Note that the mixing is confined in a thinner boundary layer for a smaller viscosity. }
  \label{fig:DifferentNu}
\end{figure}

To further explore the effects of the diffusivity and viscosity upon the mixing using the spectal method, we present figure \ref{fig:DifferentNu}.  This figure shows a sweep of viscosities (decreasing from left to right) and diffusivities (decreasing from top to bottom) which depicts the nature of the boundary layer for the passive scalar.  All of the mixing as the diffusivity and viscosity are decreased occurs in a small boundary layer adjacent to the moving wall.

\begin{figure}
  \centering
    \includegraphics[width=1\linewidth]{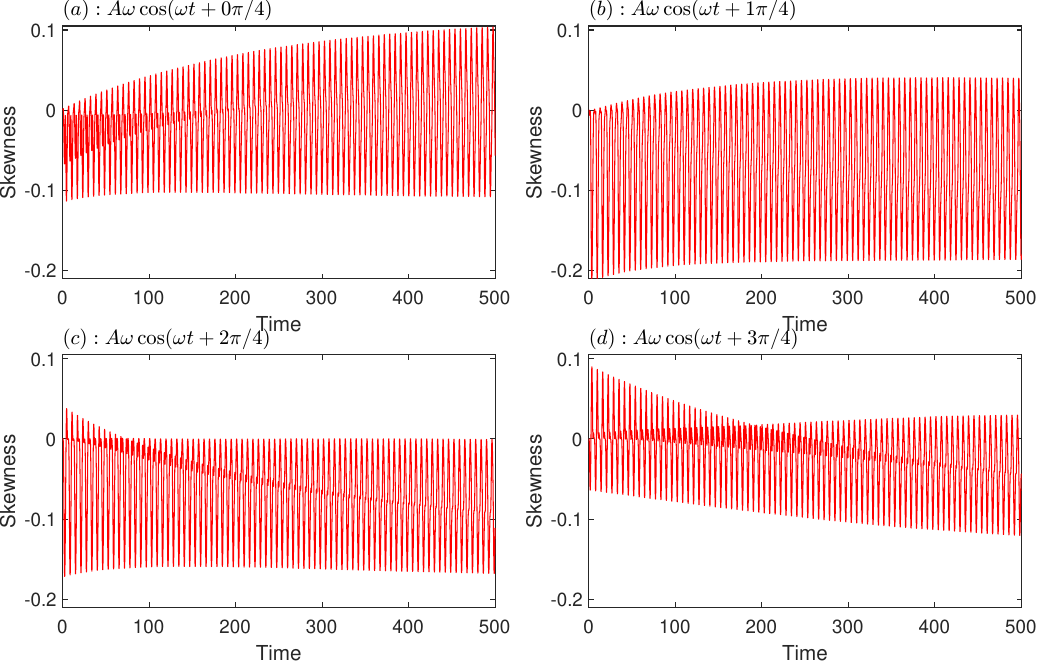}
  \caption[]
  {Skewness arising from wall velocities $A\omega\cos(\omega t+s)$ started at different phase $s$,  $(a) s=0$, $(b) s=\pi/4$, $(c) s=\pi/2$,  $(d)s=3\pi/4$, for the nonlinear Stokes layer with parameters  $\nu=0.01$ St, $\omega=0.2\pi$ rad$/s$, $L=0.2$ cm, $A=1$ cm, and $\kappa=5\times10^{-6}$ cm$^2$/s.}
  \label{fig:6}
\end{figure}

\begin{figure}
  \centering
 \includegraphics[width=0.5\linewidth]{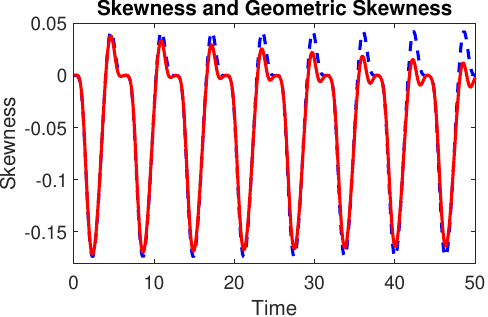}
  \caption[]
  {Comparison of short-time skewness (solid red) with analytically predicted short time asymptotic Geometric skewness (dashed blue) arising from wall motion with the velocity  $A\omega \cos(\omega t+s)$ started at phase $s=\pi/2$, for the Stokes layer solution with parameters  $\nu=0.01$ St, $\omega=0.2\pi$ rad$/s$, $L=0.2$ cm, $A=1$ cm, and $\kappa=5\times10^{-6}$ cm$^2$/s}
  \label{fig:7}
\end{figure}

We next examine the skewness behavior for a nonlinear Stokes layer with parameters $\nu=0.01$ St, $\omega=0.2\pi$ rad/s, $L=0.2$ cm, $A=1$ cm, and $\kappa=5\times10^{-6}$ cm$^2$/s and document how its sign can be controlled the initial phase of sinusoidal wall motion. The initial function is a symmetric function $T_{I}(x,y)=\left( \sqrt{2 \pi } \sigma \right)^{-1}\exp \left(-\frac{x^{2}}{2\sigma^{2}}\right) $ and $\sigma=1/40$. Shown in figure \ref{fig:6} is the evolution of the total skewness, computed using Monte-Carlo simulations, as the phase of the wall motion is changed.  Clearly the skewness shows rapid oscillation on these timescales, and the phase clearly can be used to adjust the sign of the skewness.  Lastly, in figure \ref{fig:7} we show the short time comparison of the Geometric skewness derived in the absence of diffusion with that computed with diffusion via Monte-Carlo simulations.

\section{Conclusions}
In this paper,  we develop a theory of enhanced diffusivity and skewness of the longitudinal distribution of a diffusing tracer advected by a periodic, time-varying shear flow in a straight channel. Based upon this, we present a detailed study of the tracer advected by the flows which are induced by a periodically oscillating wall in a Newtonian fluid between two infinite parallel plates as well as in an infinitely long duct.  Using a new formalism built upon the Helmholtz operator, we derive new single series formulas for the variance, effectively re-summing the double sum formulae presented in literature, e.g., Vedel et al. \cite{vedel2012transient}.  

In the study of the effective diffusion, we find the optimal  Schmidt number$f_{\mathrm{Sc}} (\omega_{0})$ or Womersley number $f_{\mathrm{Wo}} (\omega_{0})$ for mixing when the dimensionless frequency $\omega_0$ is given. The asymptotic analysis of the effective diffusivity shows that $f_{\mathrm{Sc}} (\omega_{0})\approx 2.3146$ for a large $\omega_0$ and $f_{\mathrm{Sc}} (\omega_{0})\approx \omega_{0}/6.2213$ for a small $\omega_0$. Via the relation $\omega_0=\mathrm{Wo}^{2}  \mathrm{Sc}$, we have $f_{\mathrm{Wo}} (\omega_{0})\approx \sqrt{\omega_{0}/2.3146}$ for a large $\omega_0$ and $f_{\mathrm{Wo}} (\omega_{0})\approx 2.49426 $ for a small $\omega_0$.  For fluorescein-water mixtures, we document that no interior maximum of effective diffusivity is observed because this mixture's Schmidt numbers are too large (in this case the Schmidt depends monotonically upon the temperature.  Other solute-fluid mixtures may possess enhanced diffusivities with internal maxima as a function of temperature. Further, a new mixing mechanism is identified distinguishing linear shear from the nonlinear Stokes layer.  A bound for the enhanced diffusion for the linear case is derived and shown to solely depend on the aspect ratio and molecular diffusivity, whereas for the nonlinear Stokes layer occurring at finite viscosity, the enhanced diffusion is unbounded in increasing frequency. 

In the study of the skewness, we show that the single-frequency flow can create a more symmetric distribution of the tracer than the steady flow, with skewness decay rate $t^{-3/2}$ compared to $t^{-1/2}$ for the steady case. As an extreme example, we prove the periodic time varying linear shear flow case has zero skewness for all time. Besides that, we document how the phase of the wall motion can be used to control the sign of the skewness. Experiments compare favorably with the theory and numerical simulations.  PTV flow measurements show that the experiments are well predicted by the Stokes layer solutions.  Image analysis of photographs taken at exposure times suggests that the full width at half maximum statistic is a good measure of the scalar variance and is robust to noise.   Advection-diffusion experiments with a robotically controlled moving wall show that the theory for effective diffusivity predicts the observed experimental spreading on diffusion timescales.

Future directions we intend to explore include utilizing the lubrication theory and center manifold theory \cite{mercer1990centre,beck2015analysis,beck2020rigorous} to assess the role of non-planar wall motions and their ability to further increase the effective diffusivity, along with pushing the wall motion into the stochastic regime to further understand how random wall motion creates intermittency in a passive scalar \cite{camassa2019symmetry}.

\section*{Acknowledgements}
We thank Howard A. Stone and two anonymous referees, whose comments improved the quality of the manuscript. We acknowledge funding received from the following NSF DMS-1910824 and ONR Grant No. ONR N00014-18-1-2490.

\section{Appendix}

\subsection{Multiscale analysis}\label{sec: MultiscaleAnalysis}
Following the prior work \cite{camassa2010exact},  assuming a scale separation in the initial data, we utilize multiscale analysis below to derive the effective diffusion equation induced by the periodic time-varying shear flow. We consider the following advection-diffusion equation in the parallel-plate channel with impermeable boundaries
\begin{equation}
\partial_{t} T+ u(y,t) \partial_{x} T=   \kappa \Delta T, \quad T (x,y,0) =  T_{I} \left( \frac{x}{a} \right),\quad  \left. \partial_{y} T \right|_{y=0,L}=0.
\end{equation}
We assume $\left\langle u (y,t) \right\rangle_{y,t}=0$,  where the angle bracket denotes the average of $u (y,t)$ over the region $y \times t \in [0,L] \times \mathbb{R}^{+}$. Unlike the non-dimensionalization presented in section \ref{sec:shearNondimensionalization}, here, we need two different characteristic lengths in the $x$ and $y$ direction. Hence, we introduce the following change of variables,
\begin{equation}
\begin{aligned}
&ax'=x,\quad Ly'=y, \quad \epsilon= \frac{L}{a}, \quad \frac{L^{2}}{\kappa \epsilon^2}t'=t,\quad  \frac{\kappa }{L^2}\omega'=\omega,\\
&\tilde{T}T'=T, \quad U=A\omega \quad  \mathrm{Pe}=\frac{LU}{\kappa},\quad  Uu' \left( y',\frac{t'}{\epsilon^{2}} \right)= u(y,t).  
\end{aligned}
\end{equation}
We can drop the primes without confusion and obtain the non-dimensionalized equation,
\begin{equation}
\partial_{t} T+ \frac{ \mathrm{Pe}}{\epsilon}u \left( y,\frac{t}{\epsilon^{2}} \right)\partial_{x} T=    \partial_{x}^2 T+ \frac{1}{\epsilon^{2}}\partial_{y} ^2 T,\;  T (x,y,0) =  T_{I} (x),\;  \left. \partial_{y} T \right|_{y=0,1}=0.
\end{equation}
We seek the asymptotic approximation to $T(x,y,t)$ in the limit $\epsilon\rightarrow 0$ that has the following multiscale expansion,
\begin{equation}
T(x,y,t)=T_0(x,\xi,y,t,\tau)+\epsilon T_1(x,\xi,y,t,\tau)+\epsilon^{2}T_2(x,\xi,y,t,\tau)+\mathcal{O}(\epsilon^3),
\end{equation}
with two different scales in the $x$ direction: $x$ (slow), $\xi= {x}/{\epsilon}$ (fast), and in the $t$ direction: $t$ (slow), $\tau= {t}/{\epsilon^2}$ (fast). Consequently, the differential operators along the $x$ and $t$ directions will be replaced
\begin{equation}
\begin{aligned}
& \partial_{x}\rightarrow    \partial_{x}+\frac{1}{\epsilon} \partial_{\xi}, \quad  \partial_{x}^2\rightarrow \partial_{x}^2+\frac{2}{\epsilon} \partial_{x}\partial_{\xi}+ \frac{1}{\epsilon^2} \partial_{\xi}^2,\quad \partial_{t}\rightarrow    \partial_{t}+\frac{1}{\epsilon^{2}}\partial_{\tau}.   
\end{aligned}
\end{equation}
We would have a hierarchy of equations, as one would see in a classical homogenization problem, such that the following equation holds for arbitrarily small $\epsilon$. 
For $\mathcal{O}(\epsilon^{-2})$,  we have:
\begin{equation}
\mathcal{L}T_{0}= 0, \quad
 T_0(x,\xi,y,t,\tau)|_{t=0,\tau=0}= T_{I}(x), 
\end{equation}
where  $\mathcal{L}T=\left(\partial_{\tau}+ \mathrm{Pe} u(y,\tau)\partial_{\xi}-\partial_{\xi}^2-\partial_{y}^2 \right)T$. Since the initial condition is a function of the variable $x$ only, we have $T_0(x,\xi,y,t,\tau)=T_0(x,t)
$.

For $\mathcal{O}(\epsilon^{-1})$, we have
\begin{equation}
\mathcal{L}T_1=  -\mathrm{Pe} u(y,\tau) \partial_{x} T_0+ 2\partial_{x}\partial_{\xi} T_0, \quad   T_1(x,\xi,y,0,0)= 0.
\end{equation}
The last term on the right hand side is zero. The solvability condition is guaranteed by $\left\langle  -\mathrm{Pe} u(y,\tau) \partial_{x} T_0 \right\rangle_{y,\tau}=  -\mathrm{Pe}\partial_{x} T_0 \left\langle  u(y,\tau)\right\rangle_{y,\tau}=0$. Due to the linearity of the equation, the general form of the solution is $T_1= \partial_{x} T_0(x,t) \theta(\xi,y,\tau)+C(x,t)$. Therefore, we have
\begin{equation}\label{eq:homogenization cell problem1 shear}
\mathcal{L}\theta=  -\mathrm{Pe} u,\quad \theta(\xi,y,0)=0,\quad \left. \partial_{y} \theta \right|_{y=0,1}=0.
\end{equation}
Since the initial condition and the driver are independent of $\xi$, we have $\theta(\xi, y, \tau)=\theta(y, \tau)$. For $\mathcal{O}(\epsilon^0)$, we have
\begin{equation}
\mathcal{L}T_2=  - \partial_{t} T_0-\mathrm{Pe} u(y,\tau) \partial_{x} T_1+\partial_{x}^2 T_0+ 2\partial_{x}\partial_{\xi} T_1,\: T_2(x,\xi,y,0,0)= 0.
\end{equation}
Since $\theta$ is independent of $\xi$, the last term on the right hand side is zero. The solvability condition yields the effective diffusion equation
\begin{equation}\label{eq:steady homogenization effdiffusivity}
  \partial_{t} T_0=\kappa_{\mathrm{eff}} \partial_{x}^2 T_0,\quad \kappa_{\mathrm{eff}}=1-\mathrm{Pe} \left\langle u(y, \tau) \theta \right\rangle_{y,\tau}. 
\end{equation}
Comparing equation \eqref{eq:homogenization cell problem1 shear} and \eqref{eq:Aris moment 1}, we can see that the solution $\theta$ of the cell problem is the first Aris moment. The formula of effective diffusivity \eqref{eq:steady homogenization effdiffusivity} is equivalent to equation \eqref{eq:effectiveDiffusivityDefinition}. Hence, we conclude that the Aris moment approach and the multiscale analysis approach yield the same effective diffusivity for the time-varying shear flow.  Of course, we note that the limiting procedure here, with $\epsilon \rightarrow 0$, may be different than the Aris moment approach where the limit is $t\rightarrow \infty$.

Let's use the periodic time-varying linear shear flow $u (y,t)=y \sin \omega_{0} t$ as an example. In this case, the cell problem \eqref{eq:homogenization cell problem1 shear} becomes
\begin{equation}
\partial_{\tau} \theta-\partial_{y} ^{2}\theta = - \mathrm{Pe} y \sin \omega_{0} \tau,  \quad \theta(y,0)=  0, \quad \left. \partial_{y} \theta\right|_{y=0,1}=0.
\end{equation}
The solution $\theta (y, \tau)$ has the series representation
\begin{equation}
\begin{aligned}
\theta= &\frac{\mathrm{Pe} (\cos (\tau \omega_{0} )-1)}{2 \omega_{0} }+ \frac{4 \mathrm{Pe}}{\pi^{2}} \sum\limits_{n \in \mathrm{odd}}^{}\frac{\omega_{0}  e^{-\pi ^2n^2 \tau}+\pi ^2 n^2 \sin (\tau \omega_{0} )-\omega_{0}  \cos (\tau \omega_{0} )}{ n^2 \left(\pi ^4 n^4+\omega_{0} ^2\right)} \cos n \pi y. \\
\end{aligned}
\end{equation}
Based on the formula \eqref{eq:steady homogenization effdiffusivity}, the effective diffusivity $\kappa_{\mathrm{eff}}$ is
\begin{equation}\label{eq:homogenization effdiffusivity Shear1}
\begin{aligned}
\kappa_{\mathrm{eff}}
&=1+ \frac{4 \mathrm{Pe}^{2}}{\pi^{2}}\sum\limits_{n \in \mathrm{odd}}^{}\frac{1 }{n^2 \left(\pi ^4 n^4  +\omega_{0} ^2\right)}\\
&=1+\frac{\mathrm{Pe}^{2}}{\omega_{0}^2} \left( \frac{1}{2 }-\frac{\sin \left(\frac{\sqrt{\omega_{0} }}{\sqrt{2}}\right)+\sinh \left(\frac{\sqrt{\omega_{0} }}{\sqrt{2}}\right)}{ \sqrt{2\omega_{0}} \left(\cos \left(\frac{\sqrt{\omega_{0} }}{\sqrt{2}}\right)+\cosh \left(\frac{\sqrt{\omega_{0} }}{\sqrt{2}}\right)\right)} \right),
\end{aligned}
\end{equation}
which is the same as formula \eqref{eq:effdiffusivityShear} obtained by the Aris moment approach.

\subsection{Flow equation}

\subsubsection{The Stokes wave in parallel-plate channel}\label{sec:TheStokesWaveinInfiniteChannel}
In this section, we derive the exact solution (with the transient term) of equation \eqref{eq:Stokes} and its high viscosity asymptotic expansion for completeness. The solution obtained by Laplace transform takes the form
\begin{equation}\label{eq:laplaceStokes}
  u(y,t) = \frac{1}{2 \pi \mathrm{i}} \int\limits_{C-\mathrm{i} \infty }^{ C+ \mathrm{i} \infty}e^{st}\hat{\xi}(s) \frac{\sinh \left(  \sqrt{\frac{s}{\nu}} y \right)}{ \sinh \left(  \sqrt{\frac{s}{\nu}} L \right)} \mathrm{d} s,
\end{equation}
where $\hat{\xi} (s)$ is the Laplace transform of the wall velocity $\xi(t)$. Consider a harmonic wall motion $\xi(t)=A \omega \cos \omega t$, the integrand in equation \eqref{eq:laplaceStokes} becomes
\begin{equation}
  e^{st}\hat{u}(y,s) = e^{st} \frac{As\omega }{s^2+\omega^{2}}   \frac{\sinh \left(  \sqrt{\frac{s}{\nu}} y \right)}{ \sinh \left(  \sqrt{\frac{s}{\nu}} L \right)}.
\end{equation}
The poles of $\hat{u} (y,s)$ are $s=\pm \mathrm{i} \omega$, $s=-\frac{\pi ^2 \nu  n^2}{L^2}$ for $n\in \mathbb{Z}^{+}$. By the residue theorem, we have
\begin{equation}\label{eq:2dStokesSol}
\begin{aligned}
&u(y,t)=  \text{Res}(e^{st}\hat{u}, \mathrm{i} \omega)+\mathrm{Res}(e^{st}\hat{u}, -\mathrm{i} \omega)+\sum\limits_{n=1}^{\infty}\mathrm{Res}(e^{st}\hat{u}, -\frac{\pi ^2 \nu  n^2}{L^2})  \\
&  =\Re\left(\frac{A\omega e^{\mathrm{i} \omega t} \sinh \left(e^{\mathrm{i} \frac{\pi}{4}} \frac{ \mathrm{Wo} y }{L}\right)}{\sinh \left(e^{\mathrm{i} \frac{\pi}{4}} \mathrm{Wo}\right)}\right) 
-2 \pi  A \mathrm{Wo}^{2} \sum\limits_{n=1}^{\infty} \frac{   (-1)^{-n} n  e^{-\frac{\pi ^2 \nu  n^2 t}{L^2}} \sin \left(\frac{\pi  n y}{L}\right)}{ \mathrm{Wo}^{4} +\pi ^4  n^4}.
\end{aligned}
\end{equation}
where $\mathrm{Wo}=L \sqrt{\omega / \nu}$. Since the exponential decay term will not affect the leading order of the Aris moment at long times, we neglect them in the calculation of enhanced diffusivity. 

Next, we consider the asymptotic expansion of the solution in the high viscosity limit. As $\nu \rightarrow \infty$, we have the following expansion
\begin{equation}
\frac{\sinh \left( \sqrt{ \frac{s}{\nu}} y \right)}{ \sinh \left( \sqrt{ \frac{s}{\nu}} L \right)} =
\frac{y}{L}
+\frac{L^{2}s}{6\nu} \left( \frac{y^3}{L^3}-\frac{y}{L} \right)
+ \frac{L^{4}s^{2}}{360\nu^{2}} \left( \frac{3 y^5}{L^5}-\frac{10 y^3}{L^3}+\frac{7 y}{L} \right)
+\mathcal{O}\left( \frac{L^{6}s^{3}}{\nu^{3}} \right).
\end{equation}
Then, the inverse Laplace transformation yields
\begin{equation}\label{eq:StokesWaveHighViscosity}
\begin{aligned}
 u (y,t)=&
\frac{\xi (t)y}{L}
+\frac{\xi '(t)L^{2}}{6\nu} \left( \frac{y^3}{L^3}-\frac{y}{L} \right)
+ \frac{\xi'' (t)L^{4}}{360\nu^{2}} \left( \frac{3 y^5}{L^5}-\frac{10 y^3}{L^3}+\frac{7 y}{L} \right)
+\mathcal{O} \left( \frac{L^{6}}{\nu^{3}} \right).  
\end{aligned}
\end{equation}
Particularly, for a periodic function $\xi (\omega t)$, as $\mathrm{Wo}\rightarrow 0$,  we have
\begin{equation}\label{eq:lowWoExpansion}
\begin{aligned}
 u (y,t)=&
\frac{\xi (\omega t)y}{L}
+ \frac{\xi '(\omega t) \mathrm{Wo}}{6} \left(\frac{y^3}{L^3}-\frac{y}{L} \right)\\
&+\frac{\xi'' (\omega t)\mathrm{Wo}^2 }{360}  \left(\frac{3 y^5}{L^5}-\frac{10 y^3}{L^3}+\frac{7 y}{L}\right)+\mathcal{O} \left( \mathrm{Wo}^3 \right).
  \end{aligned}
\end{equation}
For the PTV experiment presented in figure \ref{fig:PTVmesh}, the Womersley number is $\mathrm{Wo}=0.16 \sqrt{\frac{2 \pi/100 }{0.0113}} \approx 0.3773$. The low Womersley number expansion in equation \eqref{eq:lowWoExpansion} would be a good approximation for the flow in this experiment.

\subsubsection{The Stokes wave in infinite duct}
\label{sec:TheStokeswaveDuct}

In the experiment, the fluid domain is a three-dimensional space. It is natural to ask, can the Stokes layer solution derived in parallel-plate channel  approximate the Stokes layer derived in a closed duct or open duct well? We will answer this question in this section.

 In an infinitely long rectangular closed duct $y \times z \in [0,L]\times [0,H]$, the flow induced by one moving wall satisfies the equation
\begin{equation}\label{eq:3dStokesCloseEq}
\begin{aligned}
  &\partial_{t} u=  \nu \left( \partial_{y}^2 u+\partial_{z}^2 u \right),\quad u(y,z,0)=0,\\
&u(0,z,t)=0,\; u(L,z,t)=\xi( t),\; u(y,0,t)=u (y,H,t)=0.\;
\end{aligned}
\end{equation}
Applying the Laplace transform yields
\begin{equation}\label{eq:3dStokeseqLaplaceTransform}
\begin{aligned}
  & s\hat{u}=  \nu \left( \partial_{y}^2 \hat{u}+\partial_{z}^2 \hat{u} \right),\;  \hat{u}(0,z,s)=0,  \;  \hat{u}(L,z,s)=\hat{\xi}( s).
\end{aligned}
\end{equation}
For the harmonic wall motion  $\xi(t)= A \omega \cos \omega t$, we have $\hat{\xi}(s)=\frac{ A\omega s}{s^2+\omega ^2}$. According to the no-slip boundary condition at $z=0,H$, the solution takes the form
\begin{equation}\label{eq:3dStokesassump}
\begin{aligned}
&\hat{u} (y,z,s) =\sum\limits_{n=1}^{\infty} \sin \left( \frac{n z\pi }{H} \right) f_{n}(y,s).
\end{aligned}
\end{equation}
Substituting  \eqref{eq:3dStokesassump} into \eqref{eq:3dStokeseqLaplaceTransform} leads to the equation for $f_n (y,s)$
\begin{equation}
\begin{aligned}
&\left( \nu\left( \frac{n \pi }{L} \right)^{2}+s \right)f_n(y,s)=\nu \partial_{y}^2 f_{n}(y,s).\\
\end{aligned}
\end{equation}
The boundary condition $f_{n}(0,s)=0$ leads to the solution 
\begin{equation}
\begin{aligned}
f_n(y,s)=c_n\mathrm{sinh} \left( \frac{y \sqrt{H^2 s+\pi ^2 \nu  n^2}}{H \sqrt{\nu }} \right).
\end{aligned}
\end{equation} The coefficients $c_n$ can be determined by the boundary condition $\hat{u}(L,z,s)=\frac{ A\omega s}{s^2+\omega ^2}$ and the orthogonality of $\sin \left( \frac{n z\pi }{H} \right) $,
\begin{equation}
\begin{aligned}
&c_n= \frac{ 4As \omega  }{\pi  n\left( s^2+\omega ^2 \right) \mathrm{sinh} \left( \frac{L \sqrt{H^2 s+\pi ^2 \nu  n^2}}{H \sqrt{\nu }} \right) }, \quad n \in \mathrm{odd}.
\end{aligned}
\end{equation}
Hence $\hat{u} (y,z,s)$ is
\begin{equation}
\begin{aligned}
\hat{u}= & \sum\limits_{n \in \mathrm{odd}}^{\infty} \frac{ 4As \omega  \mathrm{sinh} \left( \frac{y \sqrt{H^2 s+\pi ^2 \nu  n^2}}{H \sqrt{\nu }} \right)  }{\pi  n\left( s^2+\omega ^2 \right) \mathrm{sinh} \left(\frac{L \sqrt{H^2 s+\pi ^2 \nu  n^2}}{H \sqrt{\nu }}\right) } \sin \left( \frac{n z\pi }{H} \right).\\
\end{aligned}
\end{equation}
The poles of $\hat{u} (y,z,s)$ are $s=\pm \mathrm{i} \omega$ and $s=-\frac{\pi ^2 \nu  n^2 \left(H^2+L^2\right)}{H^2 L^2}$, $n \geq 1$. By the inverse Laplace transform and residue theorem, we have the solution of equation \eqref{eq:3dStokesCloseEq}
\begin{equation}\label{eq:3dStokesCloseSol}
\begin{aligned}
&u = \frac{4 A \omega }{\pi}\sum\limits_{n \in \mathrm{odd}}^{}\Re \left( 
\frac{  e^{\mathrm{i} t \omega }\sin \left(\frac{\pi  n z}{H}\right)  \sinh \left(\frac{y \sqrt{\pi ^2 \nu  n^2+\mathrm{i} H^2 \omega }}{H \sqrt{\nu }}\right)}
{  n \text{sinh}\left(\frac{L \sqrt{\pi ^2 \nu  n^2+\mathrm{i} H^2 \omega }}{H \sqrt{\nu }}\right)}
 \right) + \mathcal{O} \left( e^{-\frac{\pi ^2 \nu  \left(H^2+L^2\right)}{H^2 L^2}} \right).\\
\end{aligned}
\end{equation}

\begin{figure}
  \centering
\includegraphics[width=1\linewidth]{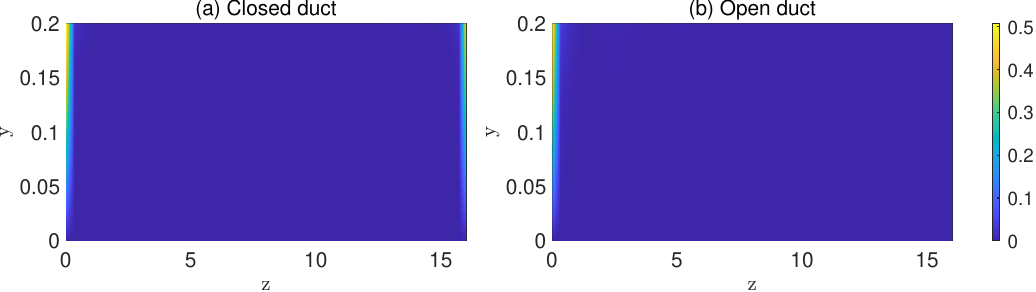}
  \caption{\textbf{Comparison of flows with different boundary conditions.} Panel (a): The difference between the solution  \eqref{eq:2dStokesSol}  in parallel-plate channel  and 105 terms of  the solution \eqref{eq:3dStokesCloseSol} in the closed duct. Panel (b): The difference between the solution in parallel-plate channel  \eqref{eq:2dStokesSol} and 105 terms of the solution \eqref{eq:3dStokesOpenSol} in the open duct. The parameters are  $\nu=0.01$St, $\omega=2\pi/100 s^{-1}$, $L=0.2$ cm, $A=1$ cm, $t=1$ s, $y\times z\in [0cm, 1/5 cm]\times [0cm, 16 cm]$ .  }
  \label{fig:Stokes2D3D}
\end{figure}

For the open duct, the no-stress boundary condition at the free surface leads to the flow equation 
\begin{equation}\label{eq:3dStokesOpenEq}
\begin{aligned}
  &\partial_{t} u=  \nu \left( \partial_{y}^2 u+ \partial_{z}^2 u \right),\; u(y,z,0)=0, \; u(0,z,t)=0,\\
& u(L,z,t)=\xi( t),\; u(y,0,t)=0, \; \left. \partial_{z} u (y,z,t) \right|_{z=H }=0.\;
\end{aligned}
\end{equation}
With the basis $\sin\left(\frac{\pi  \left(n+\frac{1}{2}\right) z}{H}\right) $, $n\geq 0$, the similar calculation yields 
\begin{equation}\label{eq:3dStokesOpenSol}
\begin{aligned}
  u= & \frac{4 A \omega  }{\pi} \sum\limits_{n=0}^{\infty}\Re \left(
    \frac{e^{\mathrm{i} t \omega } \sin \left(\frac{\pi  \left(n+\frac{1}{2}\right) z}{H}\right)  \sinh \left( \frac{y \sqrt{\nu \pi^{2} (2  n+1 )^2-4 \mathrm{i} H^2 \omega }}{2 H \sqrt{\nu }} \right)}{ (2n+1)\sinh \left( \frac{L \sqrt{\nu \pi^{2} (2  n+1 )^2-4 \mathrm{i} H^2 \omega }}{2 H \sqrt{\nu }} \right) 
    } \right)\\
  &+\mathcal{O} \left( \exp \left( -\frac{\pi ^2 \nu }{4 H^2 } \right) \right). \\
\end{aligned}
\end{equation}
Figure \ref{fig:Stokes2D3D} shows equation \eqref{eq:2dStokesSol}, \eqref{eq:3dStokesOpenSol} and \eqref{eq:3dStokesCloseSol} are only significantly different at the boundary $z=0,H$ and are indistinguishable at interior of the domain. When the tracer is concentrated at the middle of the domain, for the experimental parameters, equation  \eqref{eq:2dStokesSol} is a good approximation of \eqref{eq:3dStokesOpenSol} and \eqref{eq:3dStokesCloseSol}. 

\subsection{Lists of abbreviations}
See table \ref{tab:abbreviations}.

\begin{table}[h]
 \centering
 \begin{tabular}{l|l}
\hline
  Full Form & Abbreviation\\
\hline\hline    
Background noise subtraction  &BNS\\
Full width at half maximum &FWHM\\
Partial differential equation & PDE \\
Probability density function & PDF\\
Particle tracking velocimetry & PTV\\
Stochastic differential equation &SDE\\   
\hline
 \end{tabular}
\caption{Lists of abbreviations.} \label{tab:abbreviations}
\end{table}

\bibliographystyle{elsarticle-harv}

\end{document}